\documentclass[useAMS,usenatbib]{mn2e}

\usepackage {graphicx}
\usepackage {times}
\usepackage{pdflscape}
\usepackage{rotating}
\usepackage{natbib}
\usepackage{color}

\newcommand{\ltsima}{$\; \buildrel < \over \sim \;$}
\newcommand{\lsim}{\lower.5ex\hbox{\ltsima}}
\newcommand{\gtsima}{$\; \buildrel > \over \sim \;$}
\newcommand{\gsim}{\lower.5ex\hbox{\gtsima}}

\def\gtrsim{\mathrel{\hbox{\rlap{\hbox{\lower4pt\hbox{$\sim$}}}\hbox{$>$}}}}
\let\ga=\gtrsim
\def\lesssim{\mathrel{\hbox{\rlap{\hbox{\lower4pt\hbox{$\sim$}}}\hbox{$<$}}}}
\let\la=\lesssim

\newcommand{\nc}{\newcommand}

\nc{\be}[1]{\begin{equation}\mbox{$\label{#1}$}}
\nc{\bea}[1]{\begin{eqnarray} \mbox{$\label{#1}$}}
\nc{\Section}[2]{\section{#2}\label{#1}}
\nc{\Bibitem}[1]{\bibitem{#1}}
\nc{\Label}[1]{\label{#1}}

\nc{\vev}[1]{\langle #1 \rangle}

\nc{\eea}{\end{eqnarray}}
\nc{\ee}{\end{equation}}
\nc{\eeq}{\end{equation}}

\nc\map{{\sl WMAP\ }}

\def\simlt{\lower.5ex\hbox{\ltsima}}
\def\simgt{\lower.5ex\hbox{\gtsima}}

\def\a_sigma{ 3.1}

\title[Mid-infrared galaxy luminosity functions from AKARI]{Evolution of mid-infrared galaxy luminosity functions from the entire AKARI NEP-Deep field with new CFHT photometry
}
\author[Goto et al.]{Tomotsugu Goto$^{1}$\thanks{Based on observations obtained with MegaPrime/MegaCam and WIRCam, a joint project of CFHT, Taiwan, Korea, Canada, France, and the Canada-France-Hawaii Telescope (CFHT) which is operated by the National Research Council (NRC) of Canada, the Institute National des Sciences de l'Univers of the Centre National de la Recherche Scientifique of France, and the University of Hawaii.},
Nagisa Oi$^{2}$,
Youichi Ohyama$^{3}$,
Matthew Malkan$^{4}$,
\newauthor 
Hideo Matsuhara$^{2}$,
Takehiko Wada$^{2}$,
Marios Karouzos$^{5}$,
\newauthor 
Myungshin Im$^{5}$,
Takao Nakagawa$^{2}$,
Veronique Buat$^{6}$,
Denis Burgarella$^{6}$,
\newauthor 
Chris Sedgwick$^{7}$,
Yoshiki Toba$^{8}$,
Woong-Seob Jeong$^{9,10}$,
\newauthor 
Lucia Marchetti$^{7}$,
Katarzyna Ma\l{}ek$^{11,12}$,Ekaterina Koptelova$^{1}$,
\newauthor 
Dani Chao$^{1}$,
Yi-Han Wu$^{1}$,
Chris Pearson$^{7,13,14}$,
Toshinobu Takagi$^{2}$,
\newauthor 
Hyung Mok Lee$^{5}$,
Stephen Serjeant$^{7}$,
Tsutomu T. Takeuchi$^{11}$,
and
Seong Jin Kim$^{5,10}$
\\
$^{1}$National Tsing hua University, No. 101, Section 2, Kuang-Fu Road, Hsinchu, Taiwan 30013\\
$^{2}$Institute of Space and Astronautical Science, Japan Aerospace Exploration Agency, 	     Sagamihara, Kanagawa 252-5210\\
$^{3}$Institute of Astronomy and Astrophysics, Academia Sinica, P.O. Box 23-141, Taipei 106, Taiwan \\
$^{4}$Department of Physics and Astronomy, UCLA, Los Angeles, CA, 90095-1547, USA\\
$^{5}$Astronomy Program, Department of Physics \& Astronomy, FPRD, Seoul National University, Shillim-Dong, Kwanak-Gu, Seoul 151-742, Korea\\
$^{6}$Aix-Marseille Université, CNRS – LAM (Laboratoire d’Astrophysique de Marseille) UMR 7326, 13388 Marseille, France \\
$^{7}$Department of Physics,  The Open University, Milton Keynes, MK7 6AA, UK\\
$^{8}$ Research Center for Space and Cosmic Evolution, Ehime University, Bunkyo-cho, Matsuyama 790-8577, Japan\\
$^{9}$Korea Astronomy and Space Science Institute 61-1, Hwaam-dong, Yuseong-gu, Daejeon, Republic of Korea 305-348\\
$^{10}$Korea University of Science and Technology, 217 Gajeong-ro,
Yuseong-gu, Daejeon 305-350, Republic of Korea\\
$^{11}$Institute for Advanced Research, Nagoya University, Furo-cho, Chikusa-ku, Nagoya 464-8601\\
$^{12}$National Centre for Nuclear Research, ul. Hoza 69, 00-681 Warszawa, Poland \\
$^{13}$ RAL Space, Rutherford Appleton Laboratory, Chilton, Didcot, Oxfordshire OX11 0QX, UK\\
$^{14}$Oxford Astrophysics, Denys Wilkinson Building, University of Oxford, Keble
Rd, Oxford OX1 3RH, UK\\
}

\begin{document}

\date{\today; in original form 2015 February 20}
\pagerange{\pageref{firstpage}--\pageref{lastpage}} \pubyear{2015}
\maketitle

\label{firstpage}

\begin{abstract}
 We present infrared galaxy luminosity functions (LFs) in the AKARI North Ecliptic Pole (NEP) deep field using recently-obtained, wider CFHT optical/near-IR images.
 AKARI has obtained deep images in the mid-infrared (IR), covering 0.6 deg$^2$ of the NEP deep field.  
 However, our previous work was limited to the central area of 0.25 deg$^2$ due to the lack of optical coverage of the full AKARI NEP survey.
 To rectify the situation, we recently obtained CFHT optical and near-IR images over the entire AKARI NEP deep field. 
These new CFHT images are used to derive accurate photometric redshifts, allowing us to fully exploit the whole AKARI NEP deep field.

 AKARI's deep, continuous filter coverage in the mid-IR wavelengths (2.4, 3.2, 4.1, 7, 9, 11, 15, 18, and 24$\mu$m) exists nowhere else, due to filter gaps of other space telescopes.
It allows us to estimate restframe 8$\mu$m and 12$\mu$m luminosities without using a large extrapolation based on spectral energy distribution (SED) fitting, which was the largest uncertainty in previous studies. Total infrared luminosity (TIR) is also obtained more reliably due to the superior filter coverage.
 The resulting restframe 8$\mu$m,  12$\mu$m, and TIR LFs at $0.15<z<2.2$ are consistent with previous works, but with reduced uncertainties, especially at the high luminosity-end, thanks to the wide field coverage. 
 In terms of cosmic infrared luminosity density ($\Omega_{\mathrm{IR}}$), we found that the $\Omega_{\mathrm{IR}}$ evolves as
 $\propto (1+z)^{4.2\pm 0.4}$.
\end{abstract} 


\section{Introduction}

Studies of the extragalactic background suggest that at least half the luminous energy generated by stars has been reprocessed into the infrared (IR) emission by dust grains \citep{1999A&A...344..322L,1996A&A...308L...5P,2008A&A...487..837F}, indicating that dust-obscured star formation (SF) was  more important at higher redshifts than today.

 \citet{2005A&A...440L..17T} reported that the IR-to-UV luminosity density ratio, 
$\rho_{L(dust)}/\rho_{L(UV)}$
 evolves from 3.75 ($z$=0) to 15.1 by $z$=1.0, after a careful treatment of the sample selection effects. 
\cite{GotoTakagi2010} suggested that total infrared (TIR) luminosity accounts for $\sim$70\% of total star formation rate (SFR) density at $z$=0.25, and 90\% by $z$=1.3. 
These results highlight the importance of probing cosmic star formation activity at high redshift in the infrared bands. 

In our previous work \citep{GotoTakagi2010}, we tried to address the issue using AKARI's mid-infrared data in the NEP deep field \citep{2006PASJ...58..673M}. However, the analysis was limited to 0.25 deg$^2$, where we had deep optical imaging from Subaru/Suprime-Cam (Wada et al.  in preparation), i.e., only 40\% of the area was used for the analysis.  To improve the situation, we have recently obtained much wider optical/near-IR imaging of the entire AKARI NEP deep field using CFHT MegaCam and WIRCam. 

  AKARI's continuous mid-IR filters allow us to estimate the MIR (mid-infrared)-luminosity without using a large $k$-correction based on SED models. In this way, we are able to eliminate the largest uncertainty in previous works. 
By taking advantage of the unique filters and the larger field coverage, in this work we present more accurate estimates of restframe 8, 12$\mu$m and total infrared (TIR) LFs using the AKARI NEP-Deep data.
 We define the total IR luminosity as the integral over the rest-frame wavelength range 8-1000$\mu$m, which approximates the total luminosity emitted by interstellar dust, free from contamination by starlight.
  Unless otherwise stated, we assume a cosmology with $(h,\Omega_m,\Omega_\Lambda) = (0.7,0.3,0.7)$.

\section{Data \& Analysis}
\label{Data}

\begin{figure}
\begin{center}
\includegraphics[scale=0.6]{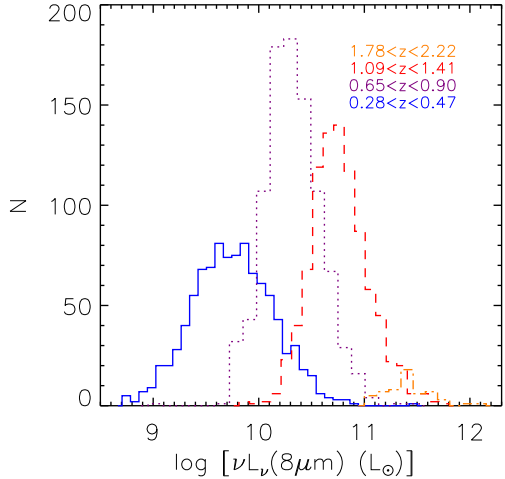}
\end{center}
\caption{
The 8$\mu$m luminosity distributions of samples used to compute  restframe  8$\mu$m LFs.
Luminosity unit is given as the logarithmic with respect to solar luminosity ($L_{\odot}$).
}\label{fig:8um_luminosity}
\end{figure}

\subsection{Multi-wavelength data in the AKARI NEP Deep field}

To obtain the TIR LFs, we adopt the method presented by \citet{GotoTakagi2010}.
 Major improvements over our previous work can be summarized as follows:
\begin{itemize}
 \item  The new AKARI NEP catalog. We have the advantage of the new AKARI 9-band mid-IR catalog in \citet{Murata2013}, which improved the corrections for scattered and stray light, artificial patterns,  and artificial sources due to bright sources. As a result, we achieved deeper 5$\sigma$ detection limits, which are 11, 9, 10, 30, 34, 57, 87, 93, and 256 $\mu$Jy in the $N2,N3,N4,S7,S9W,S11,L15,L18W$, and $L24$ filters, respectively. 
 \item Wider field coverage.
       Our previous work \citep{GotoTakagi2010} was limited to the central area of 0.25 deg$^2$, where we had deep optical data, making the results vulnerable to cosmic variance.
 We recently obtained CFHT/MegaCam $g,r,i,z, Y,J$ and, $K$ photometry (PI:Goto), which covers the entire AKARI NEP deep field (0.9 deg$^2$).  
 The wider coverage allows us to fully exploit the unique space-borne mid-IR data, while reducing the uncertainty from cosmic variance.
 \item More accurate photometric redshifts.
 Due to the availability of deeper near-IR photometry, combined with existing $u$-band data \citep{2012A&A...537A..24T}, 
 our photometric redshift accuracy was also improved to $\frac{\Delta z}{1+z}\sim$0.032 at $z<1$, and $\frac{\Delta z}{1+z}\sim$0.117 at $z>1$ \citep[see][ for details of photo-z calculation]{2014A&A...566A..60O}.
\end{itemize}

Thanks to these improvements, we have a larger sample of 5761 infrared galaxies. 
AKARI's 9 mid-IR filters are still advantageous in providing us with a unique continuous coverage at 2-24$\mu$m, where there is a gap between the Spitzer IRAC and MIPS, and the ISO $LW2$ and $LW3$.

\begin{table}
 \centering
 \begin{minipage}{180mm}
  \caption{Summary of filters used for LFs.}\label{tab:filters_used}
  \begin{tabular}{@{}ccclllcccc@{}}
  \hline
 Estimate &   Redshift & Filter \\ 
 \hline
 \hline
 8$\mu$m LF &0.28$<$z$<$0.47  &    S11 (11$\mu$m) 	\\
 8$\mu$m LF &0.65$<$z$<$0.90  &    L15 (15$\mu$m) \\
 8$\mu$m LF &1.09$<$z$<$1.41    &    L18W (18$\mu$m) \\
 8$\mu$m LF &1.78$<$z$<$2.22    &    L24 (24$\mu$m)  \\
 \hline
 12$\mu$m LF &0.15$<$z$<$0.35  &    L15 (15$\mu$m) \\
 12$\mu$m LF &0.38$<$z$<$0.62  &     L18W (18$\mu$m) \\
 12$\mu$m LF &0.84$<$z$<$1.16  &    L24 (24$\mu$m)  \\
 \hline
TIR LF & 0.2$<$z$<$0.5 &  $S7,S9W,S11,L15,L18W,$ and $L24$  \\
TIR LF &0.5$<$z$<$0.8  &  $S7,S9W,S11,L15,L18W,$ and $L24$  \\
TIR LF &0.8$<$z$<$1.2  &  $S7,S9W,S11,L15,L18W,$ and $L24$  \\
TIR LF &1.2$<$z$<$1.6  &  $S7,S9W,S11,L15,L18W,$ and $L24$  \\
 \hline
\end{tabular}
\end{minipage}
\end{table}

\subsection{Analysis}\label{sec:vmax}

We compute LFs using the 1/$V_{\max}$ method, as we did in \citet{GotoTakagi2010}.
 Uncertainties of the LF values stem from various factors such as fluctuations in 
 the number of sources in each luminosity bin, 
 the photometric redshift uncertainties,
 the $k$-correction uncertainties,
 and the flux errors. 
 To compute the errors of LFs we performed Monte Carlo simulations by creating 1000 simulated catalogs.
 Each simulated catalog contains the same number of sources, but we assigned a new redshift to each source, by following a Gaussian distribution centered at the photometric redshift with the measured dispersion $\Delta z/(1+z)$.
 The flux of each source is also changed; the new fluxes vary according to the measured flux error following a Gaussian distribution.

 For the 8$\mu$m and the 12$\mu$m LFs, 
 we can ignore the errors  due to the $k$-correction 
 thanks to the continuous AKARI MIR filter coverage. 
 The TIR errors are estimated by re-performing the SED fitting for each of the 1000 simulated catalogs (see below for more details of the SED fitting).

 We did not consider the uncertainty related to cosmic variance here since our field coverage has been significantly improved.
 For our analysis, each redshift bin covers $\sim 4\times 10^6$ Mpc$^3$ of volume.
 See \citet{2006PASJ...58..673M} for more discussion on the cosmic variance in the NEP field.

 All the other errors described above are added to the Poisson errors for each LF bin in quadrature.

\section{Results}\label{results}

\subsection{The 8$\mu$m LF}\label{sec:8umlf}

Monochromatic 8$\mu$m luminosity ($L_{8\mu m}$) is known to correlate well with the TIR luminosity \citep{2006MNRAS.370.1159B,2007ApJ...664..840H,Goto2011SDSS}, especially for star-forming galaxies, because the rest-frame 8$\mu$m flux is dominated by prominent PAH (polycyclic aromatic hydrocarbon)  features such as those at 6.2, 7.7, and 8.6 $\mu$m.
 
 Since AKARI has continuous coverage in the mid-IR wavelength range, the restframe 8$\mu$m luminosity can be obtained without a large uncertainty in $k$-correction at the corresponding redshift and filter. For example, at $z$=0.375, restframe 8$\mu$m is redshifted into $S11$ filter.
 Similarly, $L15,L18W$ and $L24$ cover restframe 8$\mu$m at $z$=0.775, 1.25 and 2.
 This filter coverage is an advantage with AKARI data. Often in previous work, SED models were used to extrapolate from Spitzer 24$\mu$m fluxes, producing the largest uncertainty. 
This is not the case for the analysis present in this paper. 
Table \ref{tab:filters_used} summarizes all filters used for our computation.

 To obtain the restframe 8$\mu$m LF, we used sources down to 80\% completeness limits in each band as measured in \citet{Murata2013}. 
We excluded those galaxies whose SEDs are better fit with QSO templates. This removed 2\% of galaxies from the sample. 

 We used the completeness curve presented in \citet{Murata2013} to correct for the incompleteness of the detections. However, this correction is 25\% at maximum, since our sample is brighter than the 80\% completeness limits. Our main conclusions are not affected by this incompleteness correction.
 To compensate for the increasing uncertainty at increasing $z$, we use four redshift bins of 0.28$<z<$0.47, 0.65$<z<$0.90, 1.09$<z<$1.41,  and 1.78$<z<$2.22.  
 We show the $L_{8\mu m}$ distribution in each redshift bin in Fig. \ref{fig:8um_luminosity}.
 Within each redshift bin, we use the 1/$V_{\max}$ method to compensate for the flux limit in each filter.

\begin{figure*}
\includegraphics[scale=0.55]{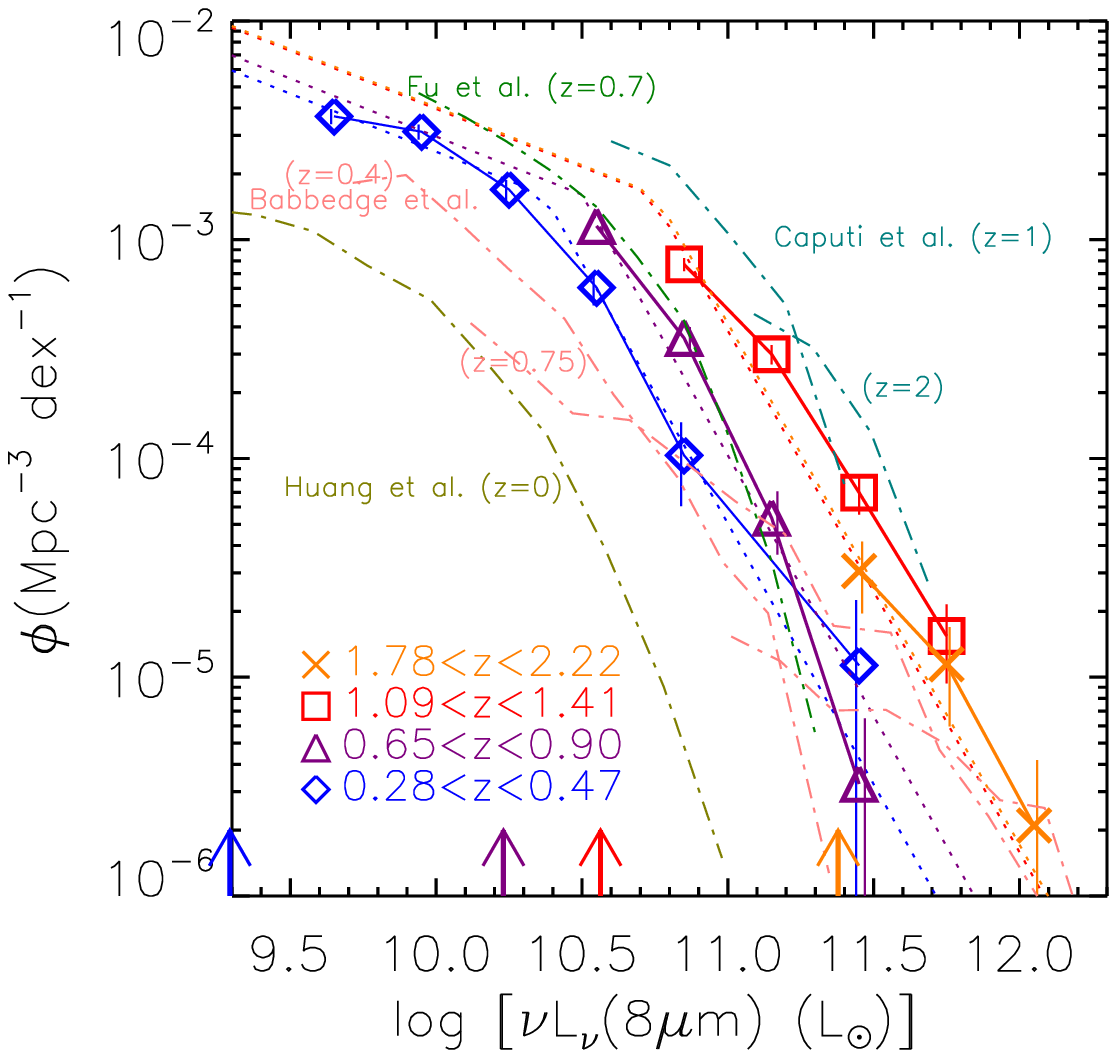}
\includegraphics[scale=0.45]{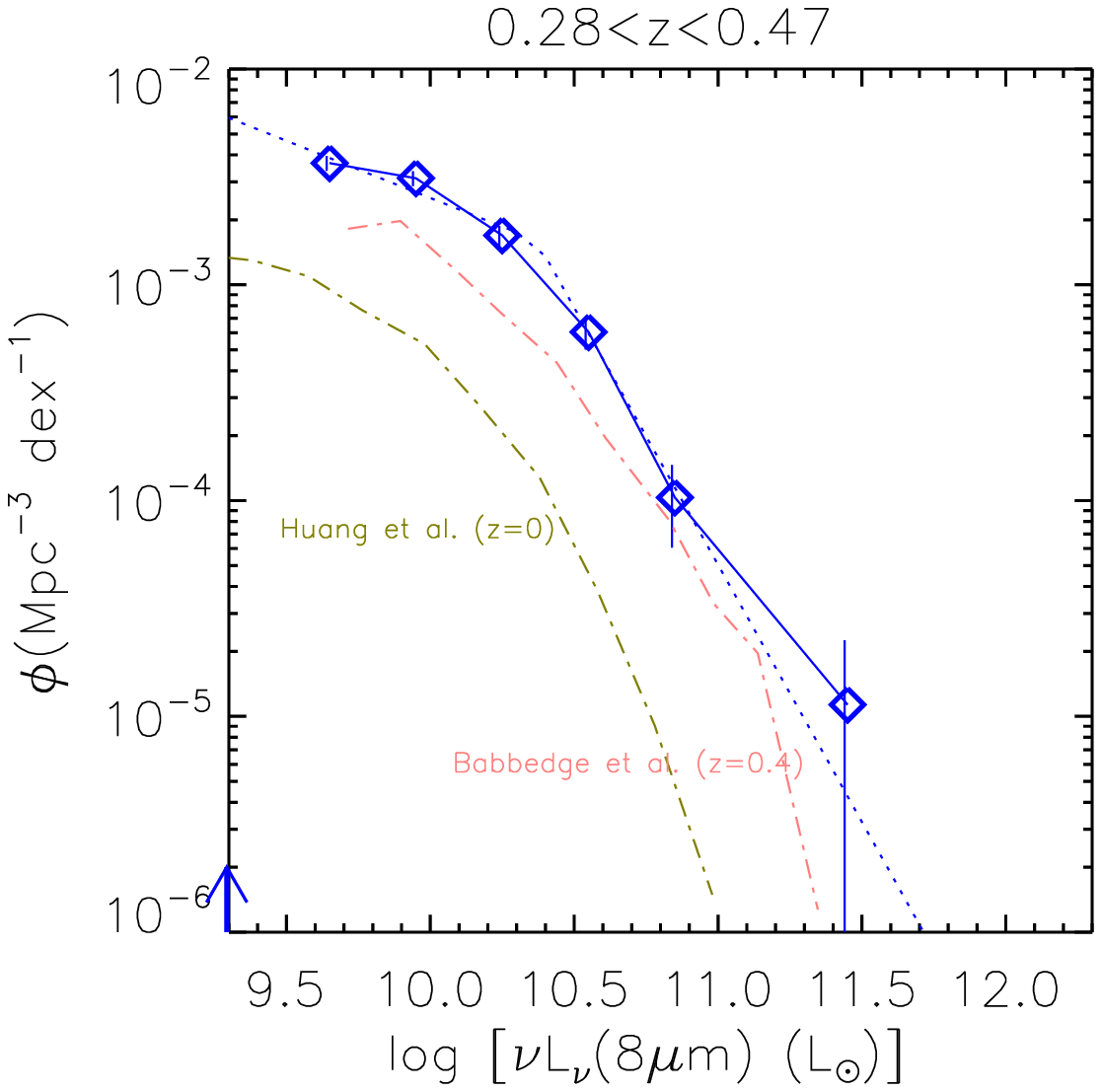}
\includegraphics[scale=0.45]{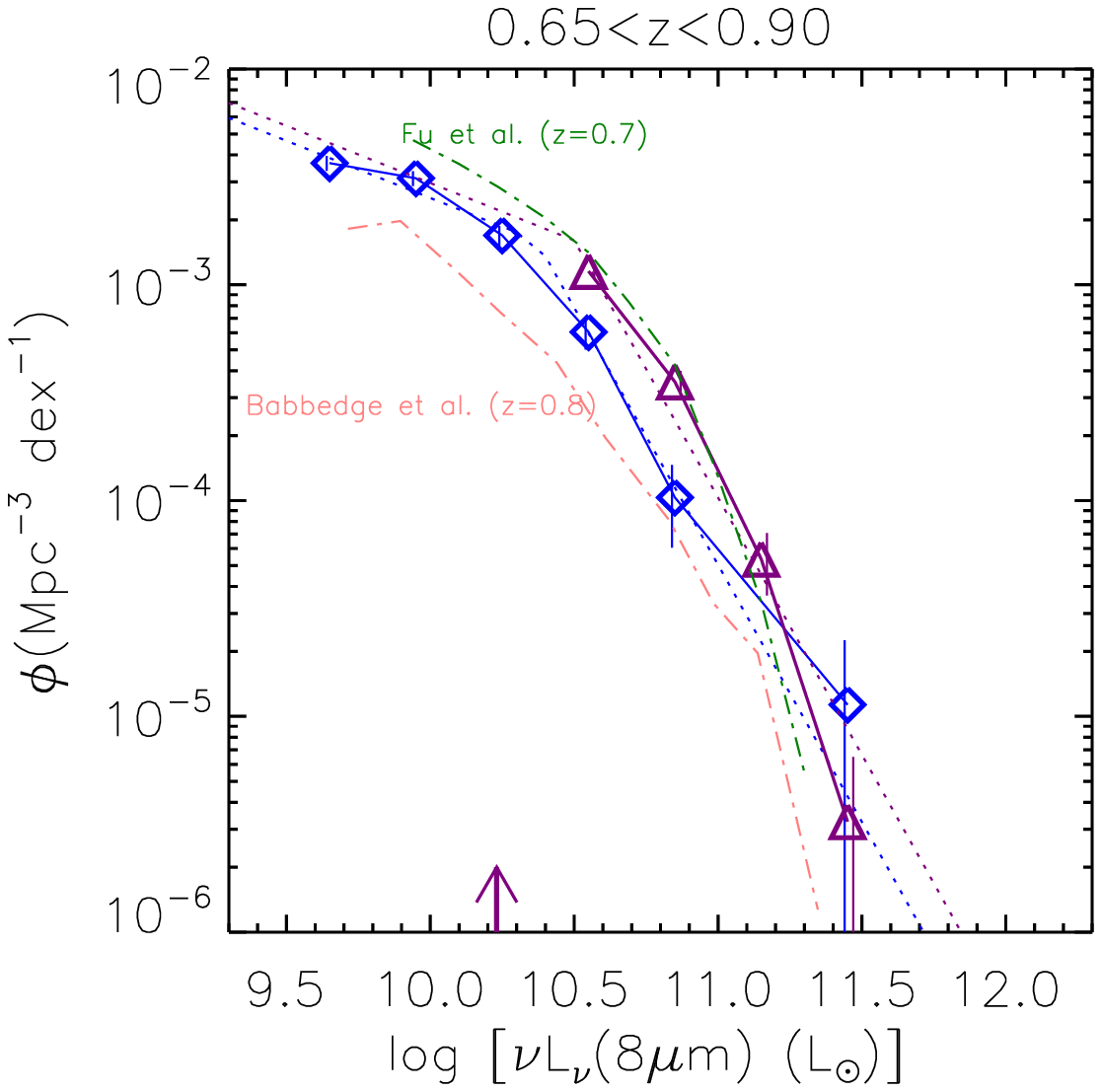}
\includegraphics[scale=0.45]{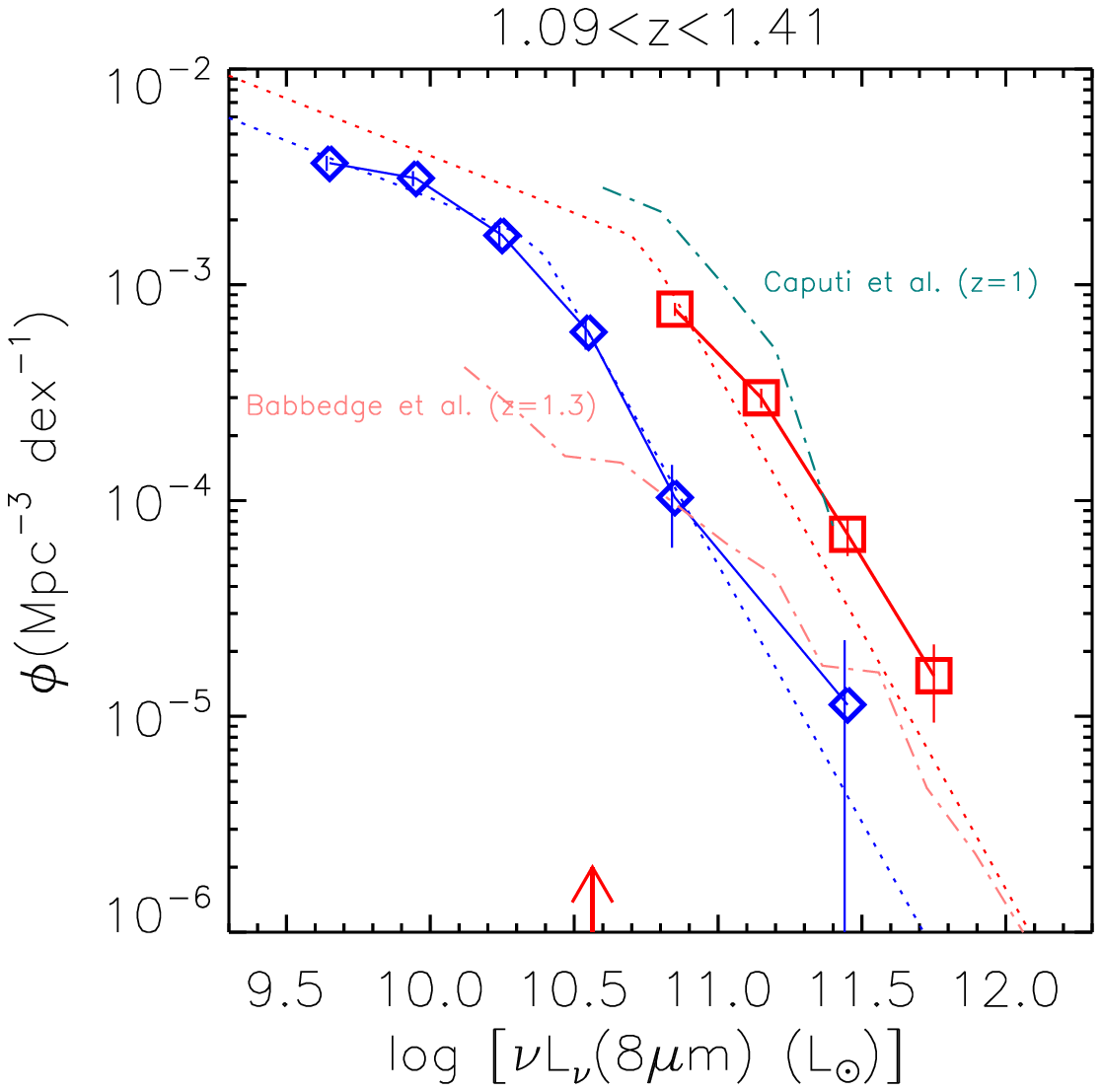}
\includegraphics[scale=0.45]{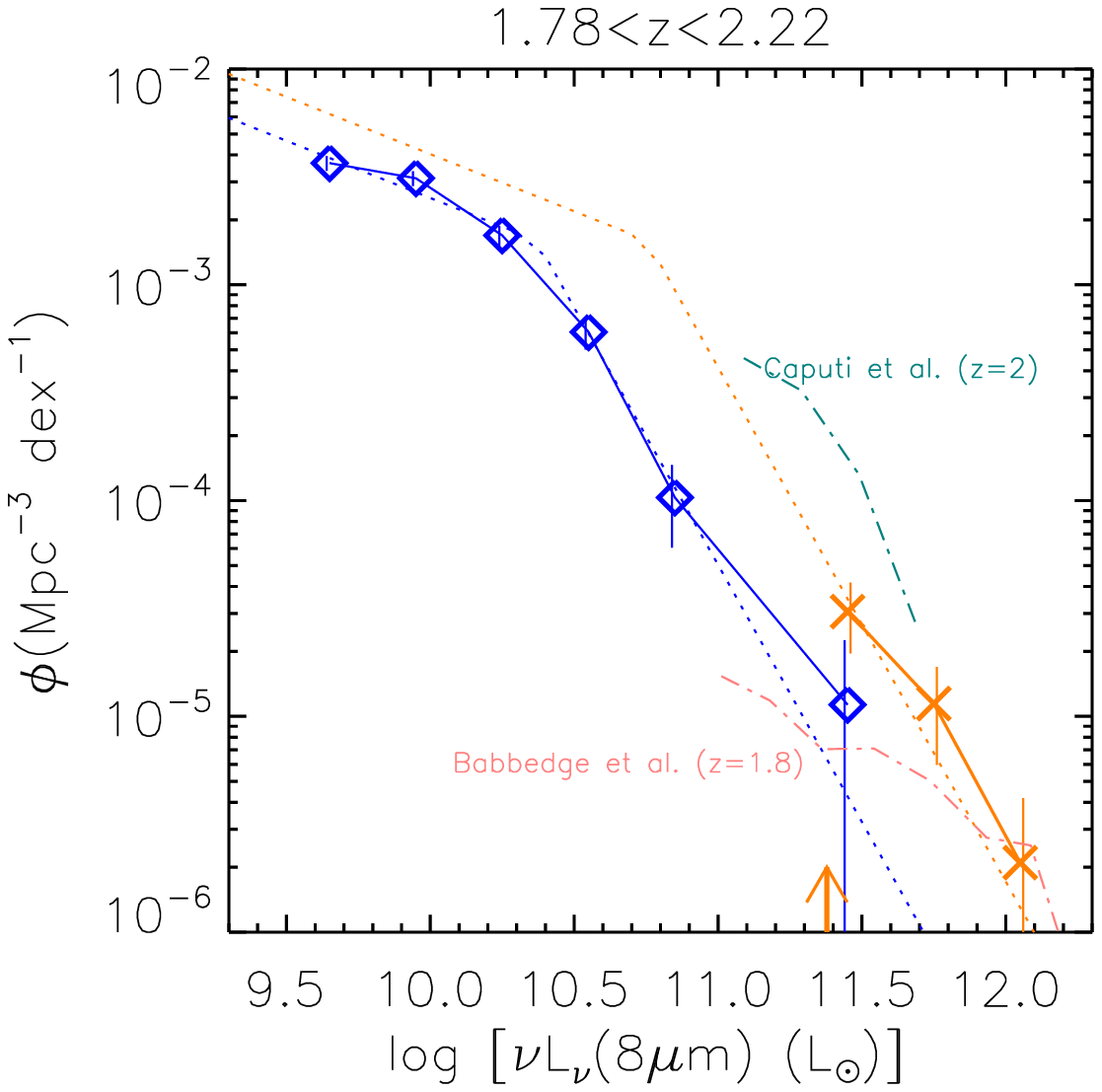}
\caption{
 Restframe  8$\mu$m LFs based on the AKARI NEP-Deep field.
The top panel shows results from all the redshift bins.
The succeeding panels show them one by one for clarity.
 The blue diamonds, the purple triangles, the red squares, and the orange crosses show the 8$\mu$m LFs at $0.28<z<0.47, 0.65<z<0.90, 1.09<z<1.41$, and $1.78<z<2.22$, respectively. AKARI's MIR filters can observe restframe 8$\mu$m at these redshifts in a corresponding filter. Error bars are estimated from the Monte Caro simulations ($\S$\ref{sec:vmax}).
 The dotted lines show analytical fits with a double-power law.
 Vertical arrows show the 8$\mu$m luminosity corresponding to the flux limit at the central redshift in each redshift bin.
 Overplotted are  \citet{2006MNRAS.370.1159B} in the pink dash-dotted lines, \citet{2007ApJ...660...97C} in the cyan dash-dotted lines, \citet{2007ApJ...664..840H} in the dark-yellow dash-dotted lines, and \citet{2010ApJ...722..653F} in the green dash-dotted line.
}\label{fig:8umlf}
\end{figure*}

 We show the computed restframe 8$\mu$m LF in Fig. \ref{fig:8umlf}. 
The arrows mark the 8$\mu$m luminosity corresponding to the flux limit at the central redshift in each redshift bin.
 Error bars on each point are based on the Monte Carlo simulation ($\S$ \ref{sec:vmax}), and are smaller than in 
 our previous work \citep{GotoTakagi2010}.
 To compare with previous work, the dark-yellow dot-dashed line also shows the 8$\mu$m LF of star-forming galaxies at $0<z<0.3$ by \citet{2007ApJ...664..840H}, 
 using the 1/$V_{\max}$  method applied to the IRAC 8$\mu$m GTO data. 
 Compared to the local LF, our  8$\mu$m LFs show strong evolution in luminosity.
 At higher redshift,  our 8$\mu$m LFs are located between those by \citet{2006MNRAS.370.1159B} and \citet{2007ApJ...660...97C}. At z=0.7, our LF is in a perfect agreement with \citet{2010ApJ...722..653F}, which is based on Spitzer IRS spectra.



\subsection{Total IR luminosity density based on the 8$\mu$m LF}\label{sec:SFR8um}
One of the important purposes in computing IR LFs is to estimate the IR luminosity density, which in turn is an extinction-free estimator of the  cosmic star formation density \citep{1998ARA&A..36..189K}. 

 We estimate  the total infrared luminosity density by integrating the LF weighted by the luminosity.
First, we need to convert $L_{8\mu m}$ to the total infrared luminosity.

A strong correlation between $L_{8\mu m}$ and total infrared luminosity ($L_{\mathrm{TIR}}$) has been reported in the literature \citep{2007ApJ...660...97C,2008A&A...479...83B}.
Using a large sample of 605 galaxies detected in the far-infrared by the AKARI all sky survey,
\citet{Goto2011IRAS} estimated 
the best-fit relation between  $L_{8\mu m}$ and  $L_{\mathrm{TIR}}$ as 
\begin{eqnarray}\label{eq:8um}
  L_{\mathrm{TIR}} = (20\pm5) \times \nu L_{\nu,8\mu m}^{0.94\pm0.01} (\pm 44\%).
\end{eqnarray}

The  $L_{\mathrm{TIR}}$ is based on AKARI's far-IR photometry in 65,90,140, and 160 $\mu$m, and the $L_{8\mu m}$ measurement is based on AKARI's 9$\mu$m photometry.
Given the superior statistics and availability over longer wavelengths (140 and 160$\mu$m),
we used this equation to convert  $L_{8\mu m}$ into $L_{\mathrm{TIR}}$.

 However, the above-mentioned conversion is based on local star-forming galaxies. It is uncertain whether it holds at higher redshift or not. 
\citet{2012ApJ...745..182N} reported that ``main-sequence'' galaxies (normal star-forming galaxies that follow the stellar-mass and SFR relation) tend to have a similar  $L_{8\mu m}$/$L_{\mathrm{TIR}}$ regardless of $L_{\mathrm{IR}}$ and redshift, up to z$\sim$2.5, and  $L_{8\mu m}$/$L_{\mathrm{TIR}}$ decreases with increasing offset above the main sequence. They suggested that this is due to a change in the ratio of polycyclic aromatic hydrocarbon (PAH) to $L_{\mathrm{TIR}}$.
On the other hand, \citet{2008A&A...479...83B} checked this by stacking 24$\mu$m sources at $1.3<z<2.3$ in the GOODS fields to find whether the stacked sources are consistent with the local relation. They concluded that the correlation is valid to link $L_{8\mu m}$ and $L_{\mathrm{TIR}}$ at  $1.3<z<2.3$.  \citet{Takagi_PAH} also showed that local $L_{7.7\mu m}$ vs $L_{\mathrm{TIR}}$ relation holds true for IR galaxies at z$\sim$1 (see their Fig.10). \citet{2008ApJ...675.1171P} showed that $z\sim$2 sub-millimeter galaxies lie on the relation between $L_{\mathrm{TIR}}$  and $L_{PAH,7.7}$ that has been established for local starburst galaxies. 
The $S_{70}/S_{24}$ ratios of 70$\mu$m sources in \citet{2007ApJ...668...45P} are also consistent with the local SED templates. Following these results, we use equation (\ref{eq:8um}) for our sample. However, we should keep in mind that there may be a possible evolution with redshift as discussed in \citet{2008ApJ...675..262R,2009ApJ...700..183H}.
 
The conversion, however, has been the largest source of error in estimating  $L_{\mathrm{TIR}}$ values from  $L_{8\mu m}$.  Reported dispersions are 37, 55 and 44\% by \citet{2008A&A...479...83B}, \citet{2007ApJ...660...97C}, and \citet{Goto2011IRAS}, respectively.  It should be kept in mind that the restframe $8\mu$m is sensitive to the star-formation activity, but at the same time, it is where the SED models have strongest discrepancies due to the complicated PAH emission lines. Possible SED evolution, and the presence of (unremoved) AGN will induce further uncertainty. A detailed comparison of different conversions is presented in Fig.12 of  \citet{2007ApJ...660...97C}, who reported a factor of $\sim$5 differences among various models.


Having these cautions in mind, the 8$\mu$m LF is weighted by the $L_{\mathrm{TIR}}$ and integrated to obtain the TIR density.
For integration, we first fitted an analytical function to the LFs.
In the literature, IR LFs were fit better by a double-power law \citep{2006MNRAS.370.1159B} or a double-exponential \citep{1990MNRAS.242..318S,2004ApJ...609..122P,2006A&A...448..525T,2005ApJ...632..169L} than a Schechter function, which declines too steeply at high luminosities, underestimating
the number of bright galaxies.  
In this work, we fit the 8$\mu$m LFs using a double-power law \citep{2006MNRAS.370.1159B} as is shown below.

\begin{equation}
 \label{eqn:lumfunc2p}
 \Phi(L)dL/L^{*} = \Phi^{*}\bigg(\frac {L}{L^{*}}\bigg)^{1-\alpha}dL/L^{*}, ~~~ (L<L^{*})
\end{equation}

\begin{equation}
 \label{eqn:lumfunc2p2}
 \Phi(L)dL/L^{*} = \Phi^{*}\bigg(\frac {L}{L^{*}}\bigg)^{1-\beta}dL/L^{*}, ~~~  (L>L^{*})
\end{equation}

\noindent First, the  double-power law is fitted to the lowest redshift LF at 0.28$<z<$0.47 to determine the normalization ($\Phi^{*}$) and slopes ($\alpha,\beta$). 
 For higher redshifts we do not have enough statistics to simultaneously fit 4 parameters ($\Phi^{*}$, $L^*$, $\alpha$, and $\beta$).  Therefore, we fixed the slopes and normalization at the local values and varied only $L^*$ for the higher-redshift LFs.
 Fixing the faint-end slope is a common procedure with the depth of current IR satellite surveys \citep{2006MNRAS.370.1159B,2007ApJ...660...97C}.
 The stronger evolution in luminosity than in density found by previous work \citep{2005ApJ...630...82P,2005ApJ...632..169L} also justifies this parametrization. 
 The best-fit parameters are presented in Table \ref{tab:fit_parameters}.
  We found that our $L^*_{8\mu m}$ evolves as $\propto (1+z)^{1.4\pm0.7}$ in the range of $0.48<z<2$.
 Once the best-fit parameters are found, we integrate the double power law outside the luminosity range in which we have data to estimate the TIR luminosity density, $\Omega_{\mathrm{TIR}}$.

\begin{table*}
 \centering
 \begin{minipage}{180mm}
  \caption{Best fit parameters for 8,12$\mu$m and TIR LFs}\label{tab:fit_parameters}
  \begin{tabular}{@{}ccccllcccc@{}}
  \hline
   Redshift & LF & $L^*$ ($ 10^{10} L_{\odot}$)& $\Phi^*(10^{-3} \mathrm{Mpc^{-3} dex^{-1}})$ & $\alpha$ & $\beta$  \\ 
 \hline
 \hline
0.28$<$z$<$0.47  &   8$\mu$m & $2.3^{+0.2}_{-0.2}$ & 1.6$^{+0.2}_{-0.3}$ & 1.53$^{+0.01}_{-0.02}$   &   3.4$^{+0.6}_{-0.5}$ 	\\
0.65$<$z$<$0.90  &   8$\mu$m & $3.1^{+0.1}_{-0.1} $ & $1.6 $  &1.53 & 3.4	\\
1.09$<$z$<$1.41    &   8$\mu$m & $5.5^{+0.1}_{-0.2} $ & $1.6$  &1.53 & 3.4	\\
1.78$<$z$<$2.22    &   8$\mu$m & $5.7^{+0.8}_{-1.1} $ & $1.6$  &1.53 & 3.4	\\
 \hline

0.15$<$z$<$0.35  &   12$\mu$m & $1.1^{+0.2}_{-0.1} $ &  1.1$^{+0.2}_{-0.3}$ &1.91$^{+0.05}_{-0.04}$   &   3.4$^{+1.4}_{-0.4}$ \\
0.38$<$z$<$0.62  &   12$\mu$m & 2.4$^{+0.2}_{-0.2} $ &  1.1 &1.91   &  3.4 \\
0.84$<$z$<$1.16  &   12$\mu$m & 4.6$^{+0.3}_{-0.3} $ &  1.1 &1.91   &   3.4 \\
 \hline
0.2$<$z$<$0.5  &  Total  & 10$^{+1}_{-1} $& 3.4$^{+0.2}_{-0.1}$ &  1.4$^{+0.1}_{-0.1}$   &   3.3$^{+0.2}_{-0.1}$ 	\\
0.5$<$z$<$0.8  &  Total  & 18$^{+1}_{-2} $ &3.4 &  1.4   &   3.3	\\
0.8$<$z$<$1.2  &  Total  & 31$^{+3}_{-2} $ &3.4 &  1.4   &   3.3	\\
1.2$<$z$<$1.6  &  Total  & 140$^{+10}_{-22} $   &  3.4 &  1.4   &   3.3	\\
 \hline
%
\end{tabular}
\end{minipage}
\end{table*}

The final resulting total luminosity density ($\Omega_{\mathrm{IR}}$) is shown in  $\S$\ref{sec:SFH} along with results obtained from other wavelengths.

\subsection{12$\mu$m LF}\label{sec:12umlf}

\begin{figure}
\begin{center}
\includegraphics[scale=0.6]{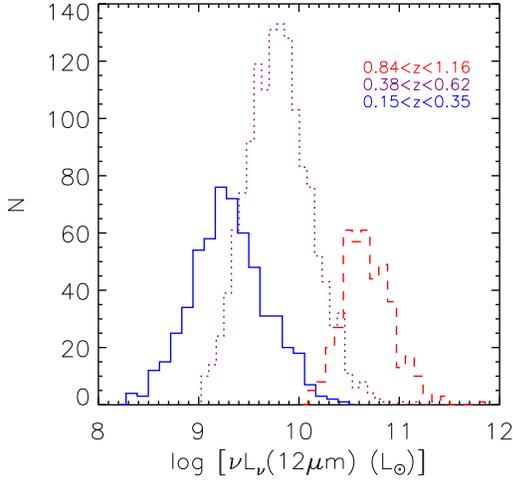}
\end{center}
\caption{
The 12$\mu$m luminosity distributions of samples used to compute restframe  12$\mu$m LFs. 
The Luminosity unit is in logarithmic solar units ($L_{\odot}$).
}\label{fig:12um_luminosity}
\end{figure}

The 12$\mu$m luminosity ($L_{12\mu m}$) has been well studied through ISO and IRAS. It is known to correlate closely with the TIR luminosity \citep{1995ApJ...453..616S,2005ApJ...630...82P}. 
 
 As was the case for the 8$\mu$m LF, it is advantageous that AKARI's continuous filters in the mid-IR allow us to estimate restframe 12$\mu$m luminosity without much extrapolation based on SED models.
 
\begin{figure*}
\begin{center}
\includegraphics[scale=0.45]{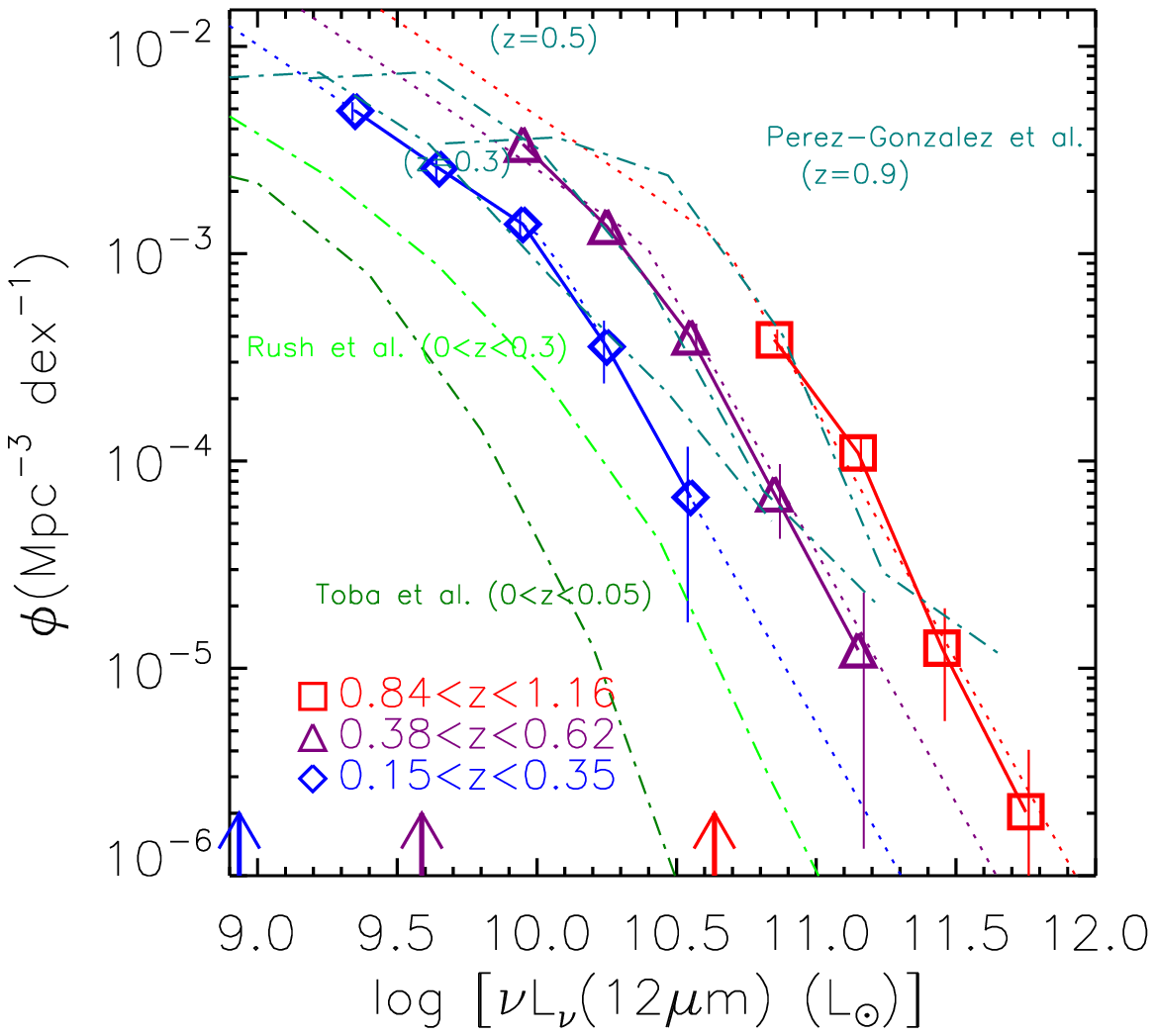}
\includegraphics[scale=0.45]{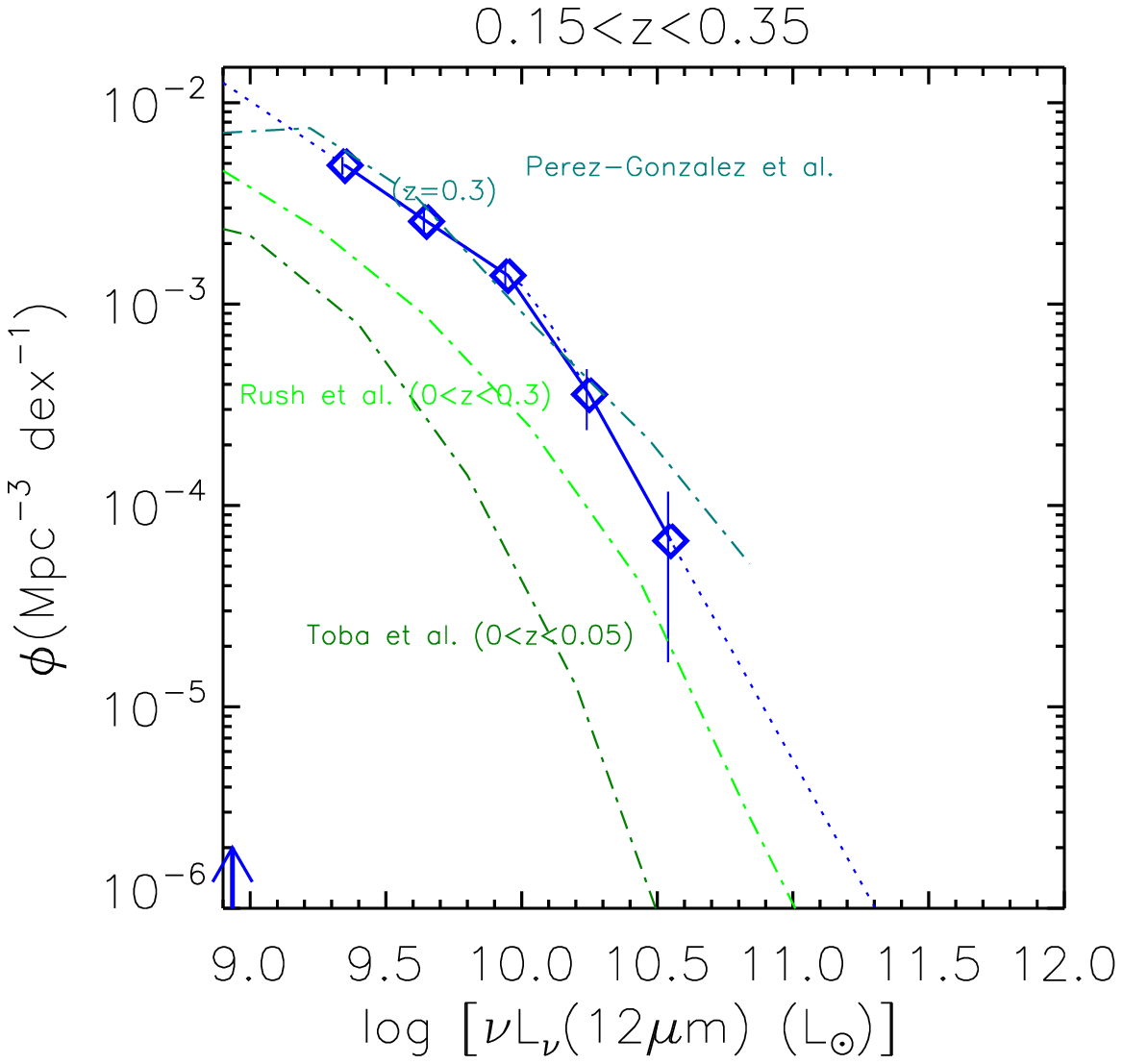}
\includegraphics[scale=0.45]{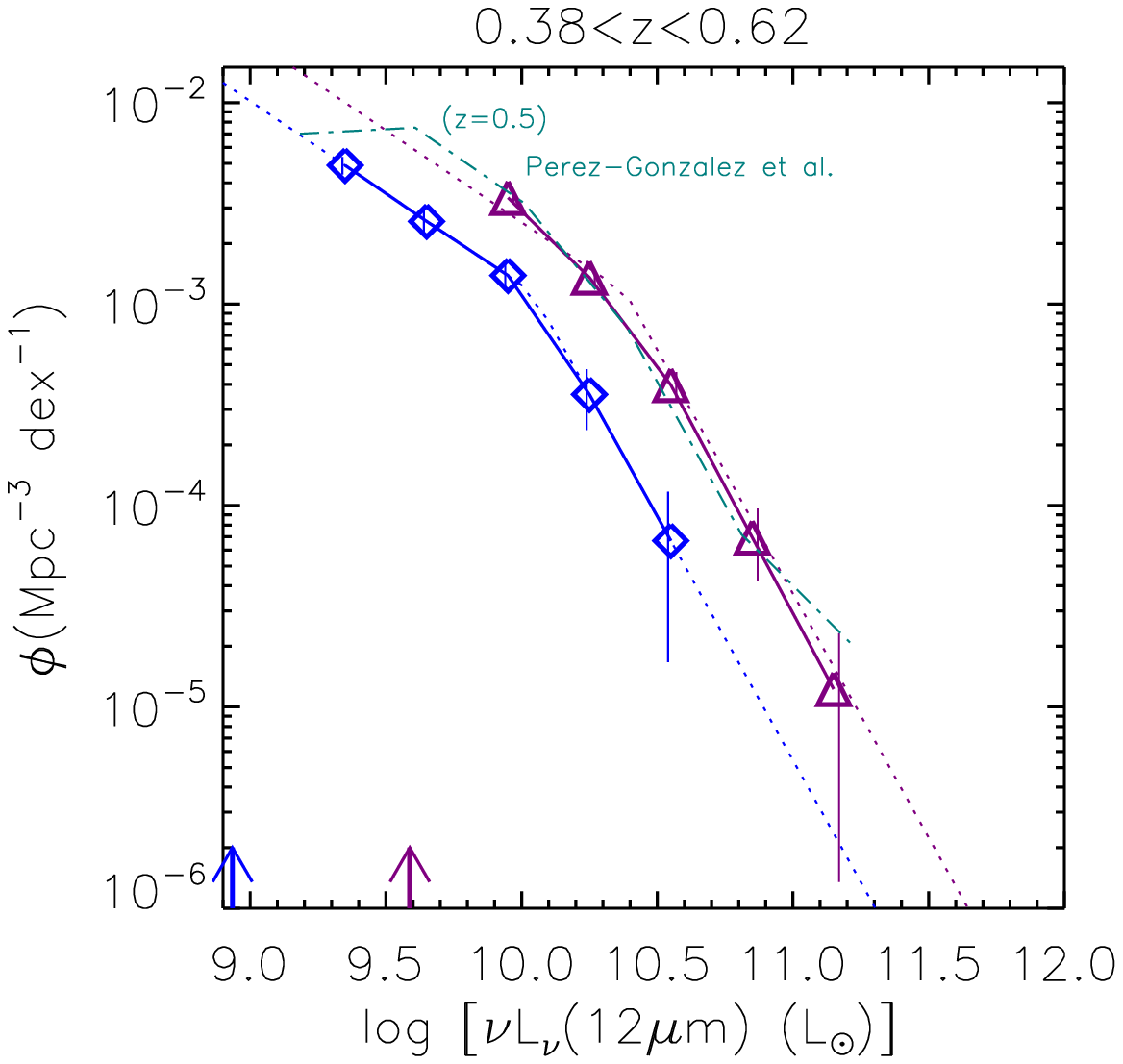}
\includegraphics[scale=0.45]{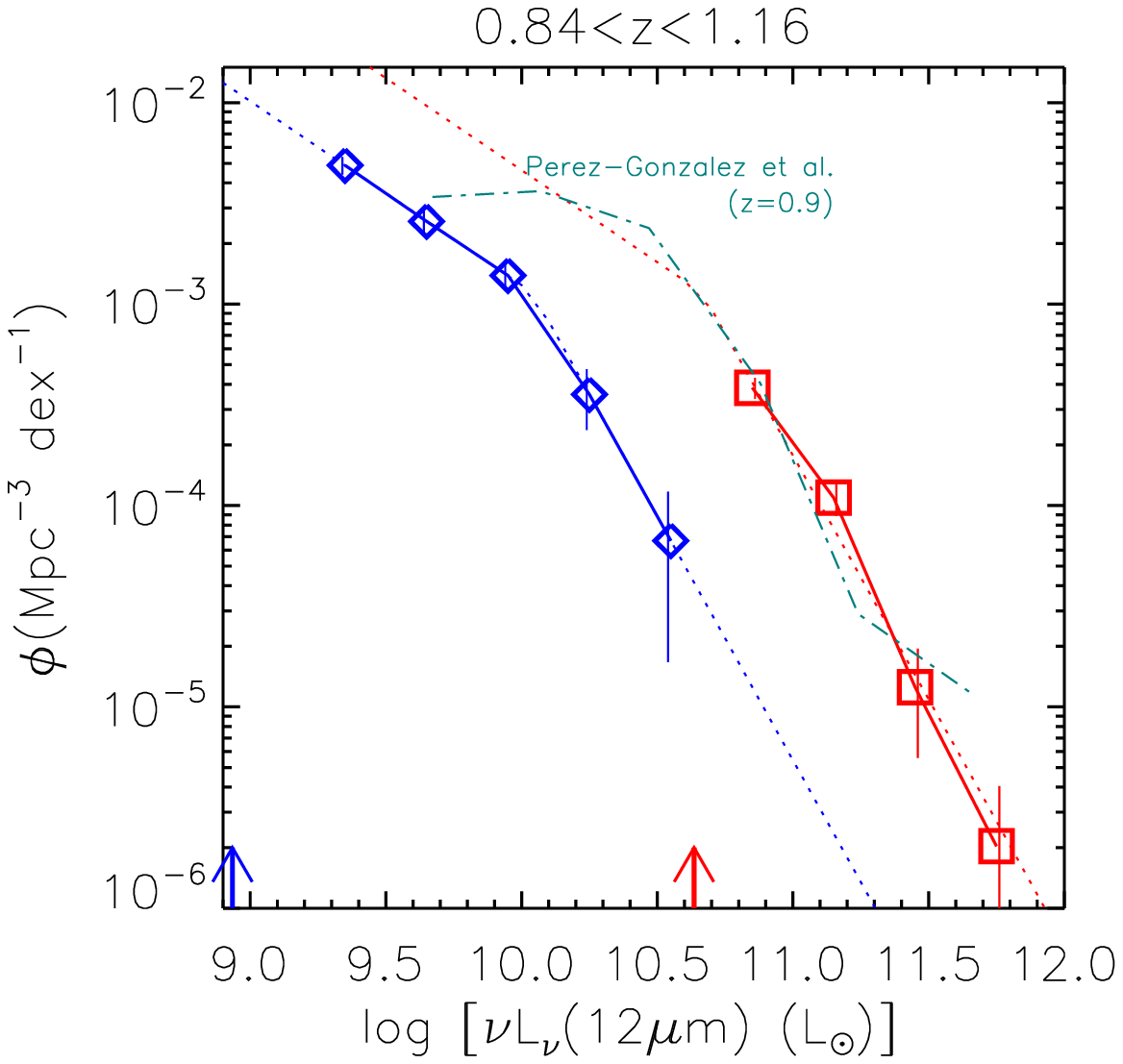}
\end{center}
\caption{
Restframe  12$\mu$m LFs based on the AKARI NEP-Deep field.
Luminosity unit is logarithmic solar luminosity ($L_{\odot}$).
  The blue diamonds, the purple triangles, and the red squares show the 12$\mu$m LFs at $0.15<z<0.35, 0.38<z<0.62$, and $0.84<z<1.16$, respectively.
In the upper-left panel, we plot results from redshifts together. In the succeeding panels, we compare results from different redshifts separately for clarity. 
  Vertical arrows show the 12$\mu$m luminosity corresponding to the flux limit at the central redshift in each redshift bin.
  Overplotted are  \citet{2005ApJ...630...82P} at $z$=0.3, 0.5 and 0.9 in the dark-cyan dash-dotted lines, 
\citet{2014ApJ...788...45T} at 0$<z<$0.05 based on WISE in the dark green dash-dotted lines,
and \citet{1993ApJS...89....1R} at 0$<z<$0.3 in the light green dash-dotted lines. Note \citet{1993ApJS...89....1R} is at higher redshifts than \citet{2014ApJ...788...45T}.
}\label{fig:12umlf}
\end{figure*}

At targeted redshifts of $z$=0.25, 0.5, and 1, the $L15,L18W$ and $L24$ filters cover the restframe 12$\mu$m, respectively.
 We summarize the filters used in Table \ref{tab:filters_used}.
 The methodology is the same as for the 8$\mu$m LF; we used the sample down to the 80\% completeness limit, corrected for the incompleteness, then used the 1/$V_{\max}$  method to compute the LF in each redshift bin.
The $L_{12\mu m}$ distribution is presented in Fig. \ref{fig:12um_luminosity}.
 The resulting 12$\mu$m LF is shown in Fig. \ref{fig:12umlf}.
 The light green dash-dotted line shows   12$\mu$m  LF based on 893 galaxies at $0<z<0.3$ in the IRAS Faint Source Catalog \citep{1993ApJS...89....1R}. 
 The dark green dash-dotted line shows 12$\mu$m  LF at $0.006<z<0.05$ based on 223,982 galaxies from WISE sources in Table 7 of \citet{2014ApJ...788...45T}.
 Compared with these $z$=0 LFs, the 12$\mu$m LFs show steady evolution with increasing redshift. 
 In the range of $0.25<z<1$, the $L^*_{12\mu m}$ evolves as $\propto (1+z)^{1.5\pm0.4}$.

\subsection{Total IR luminosity density based on the 12$\mu$m LF}\label{sec:SFR12um}

 The 12$\mu$m is one of the most frequently used monochromatic fluxes to estimate $L_{\mathrm{TIR}}$.
 The total infrared luminosity can be computed from the $L_{12\mu m}$ using the conversion in \citet{2001ApJ...556..562C,2005ApJ...630...82P}.

\begin{eqnarray}\label{eq:12um}
\log L_{\mathrm{TIR}}=\log (0.89^{+0.38}_{-0.27})+1.094 \log L_{12\mu m}\label{Aug  7 17:32:13 2009}
\end{eqnarray}

\citet{2005A&A...432..423T} independently estimated the relation to be 

\begin{eqnarray}\label{eq:12um_takeuchi}
\log L_{\mathrm{TIR}}=1.02+0.972 \log L_{12\mu m},
\end{eqnarray}

 which we also use to check our $L_{\mathrm{TIR}}-L_{12\mu m}$ conversion. As both authors state, these conversions contain an error of a factor of 2-3. Therefore, we should avoid conclusions that could be affected by such errors. 
The computed total luminosity density based on the 12$\mu$m LF using equation (\ref{eq:12um_takeuchi}) will be presented in  $\S$\ref{sec:SFH}.

\subsection{TIR LF} \label{sec:tirlf}

AKARI's continuous mid-IR coverage is also superior for SED-fitting to estimate $L_{\mathrm{TIR}}$. 
This is because for star-forming galaxies, the mid-IR part of the IR SED is dominated by the PAH emission lines,
 which reflect the SFR of galaxies \citep{1998ApJ...498..579G}, and thus, correlates well with $L_{\mathrm{TIR}}$, which is also a good indicator of the galaxy SFR. 

After photometric redshifts are estimated using the UV-optical-NIR photometry, we fix the redshift at the photo-$z$ \citep{2014A&A...566A..60O}, then use the same {\ttfamily  LePhare} code to fit the infrared part of the SED to estimate TIR luminosity. 
We used \citet{2003MNRAS.338..555L}'s SED templates to fit the photometry using the AKARI bands at $>$6$\mu$m ($S7,S9W,S11,L15,L18W$ and $L24$). 

In the mid-IR, color-correction could be large when strong PAH emissions shift into the bandpass (a factor of $\sim$3). However, during the SED fitting, we integrate the flux over the bandpass weighted by the response function. Therefore, we do not use the flux at a fixed wavelength. As such, the color-correction is negligible in our process (a few percent at most).
 We also checked that using different SED models \citep{2001ApJ...556..562C,2002ApJ...576..159D}  does not change our essential results.

 Galaxies in the targeted redshift range are best sampled in the 18$\mu$m band due to the wide bandpass of the $L18W$ filter \citep{2006PASJ...58..673M}. In fact, in a single-band detection, the 18$\mu$m image returns the largest number of sources.
 Therefore, we applied the 1/$V_{\max}$  method using the detection limit at $L18W$.
 We also checked that using the $L15$ flux limit does not change our main results.
 The same \citet{2003MNRAS.338..555L}'s models are also used for $k$-corrections necessary to compute $V_{\max}$  and $V_{\min}$.
 The redshift bins used are 0.2$<z<$0.5, 0.5$<z<$0.8, 0.8$<z<$1.2,  and 1.2$<z<$1.6.  

\begin{figure*}
\begin{center}
\includegraphics[scale=0.55]{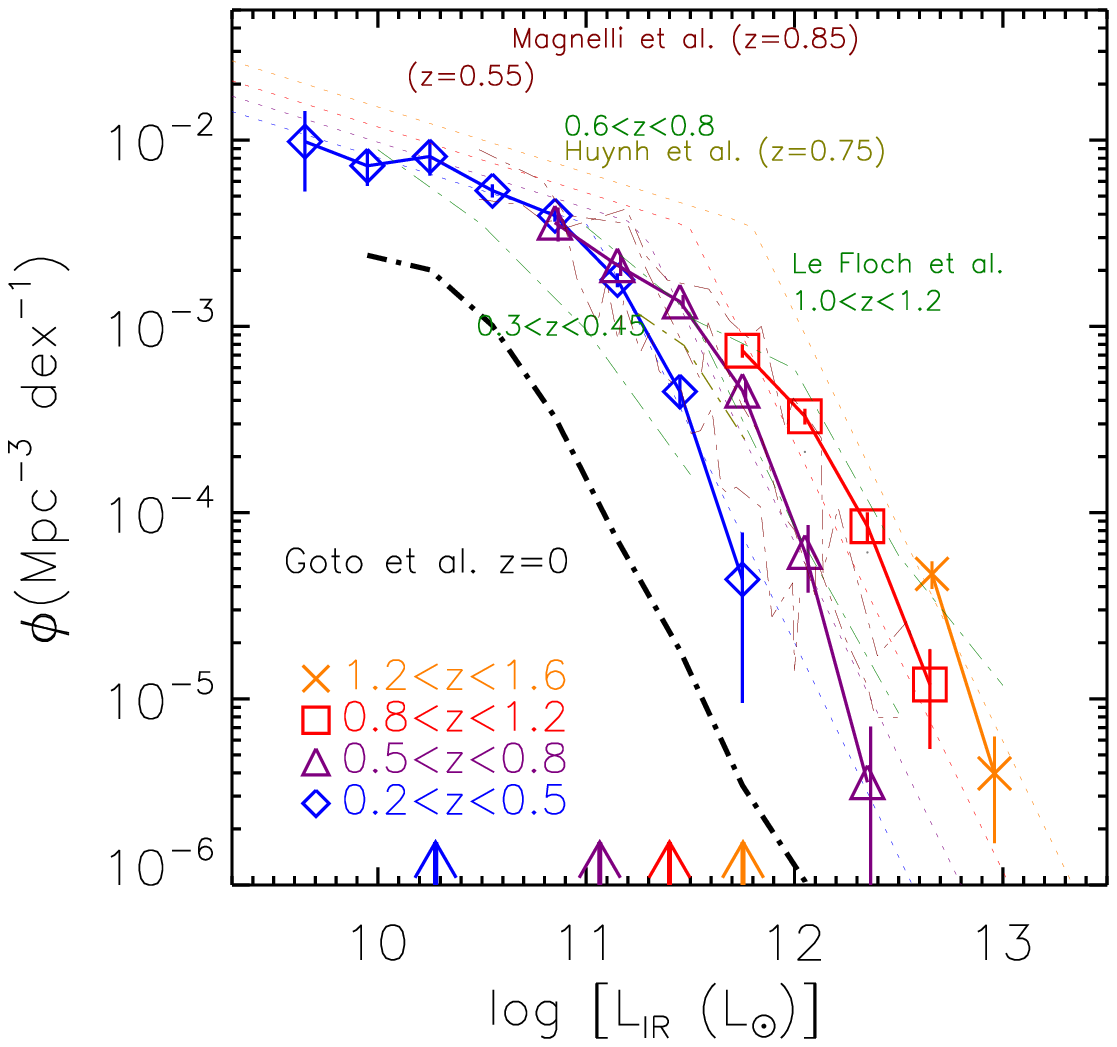}
\includegraphics[scale=0.45]{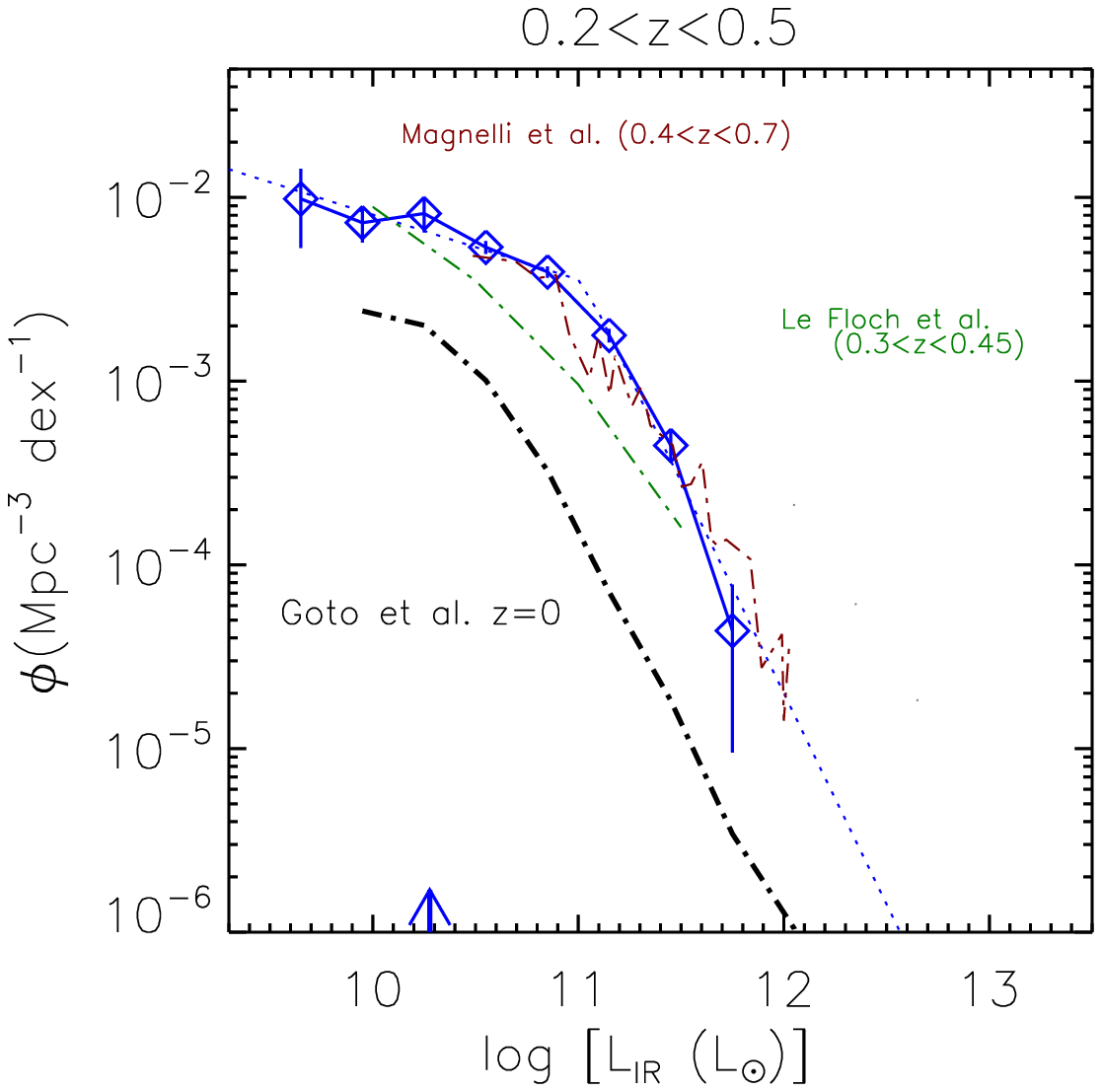}
\includegraphics[scale=.45]{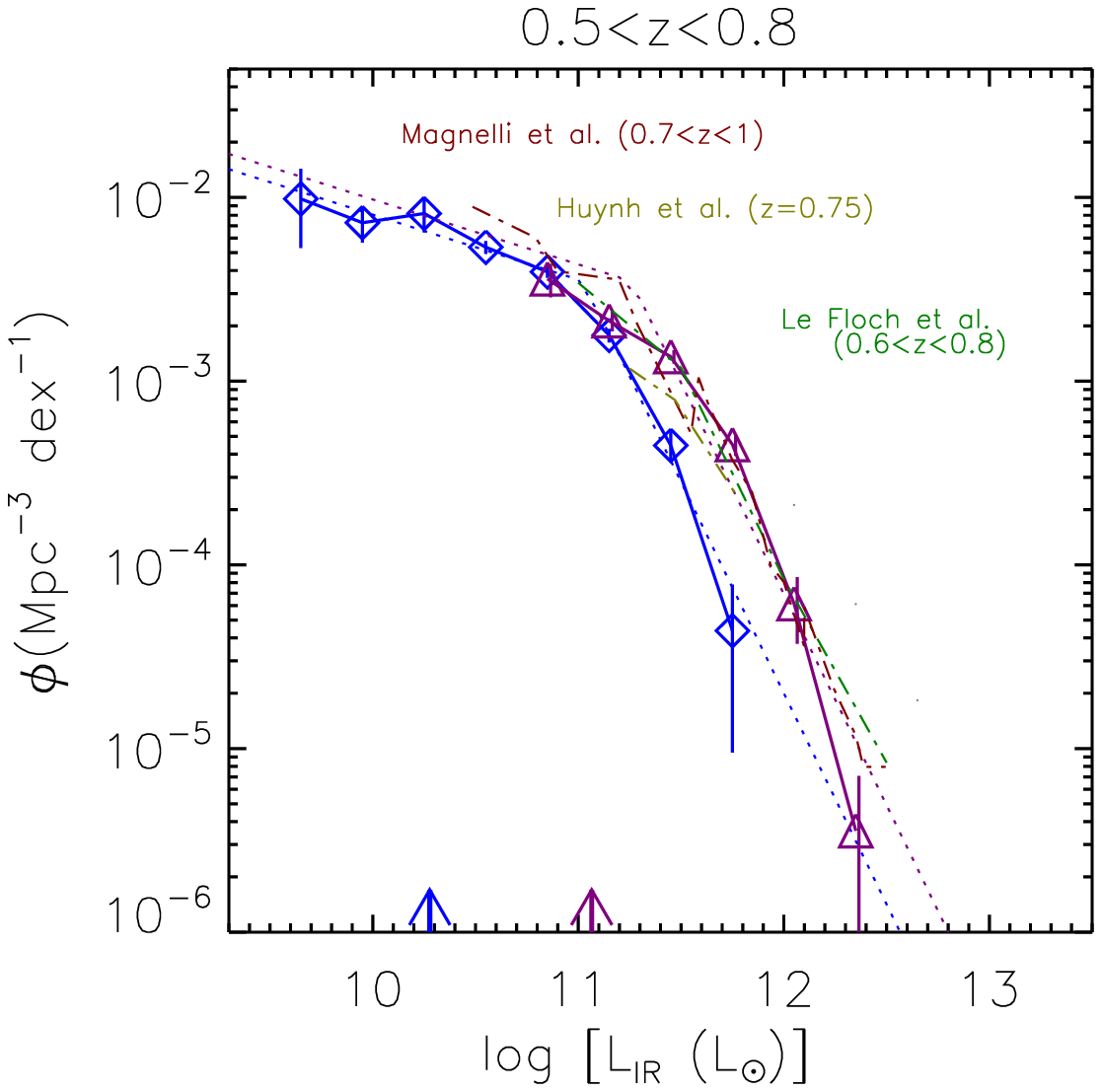}
\includegraphics[scale=.45]{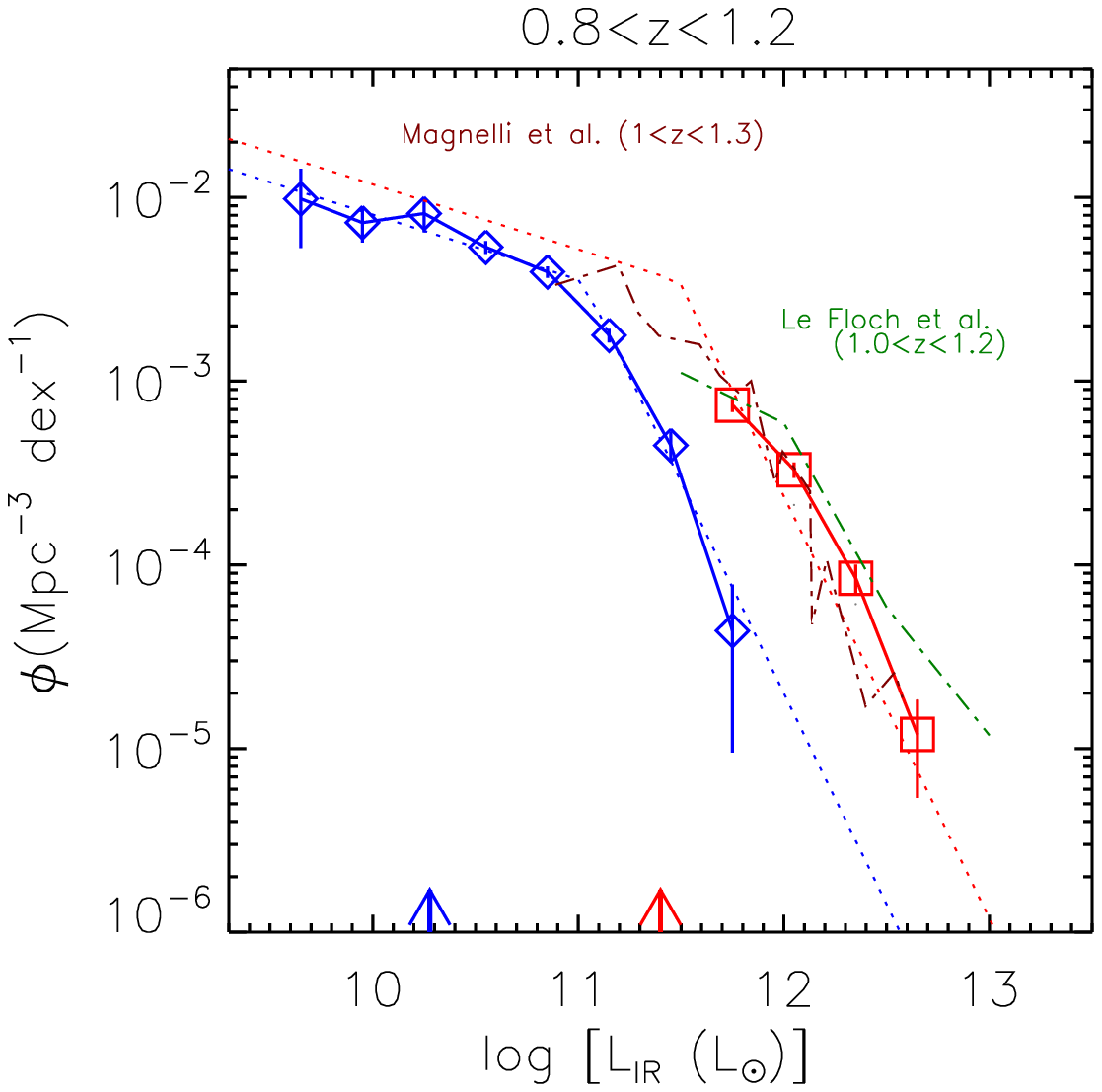}
\includegraphics[scale=.45]{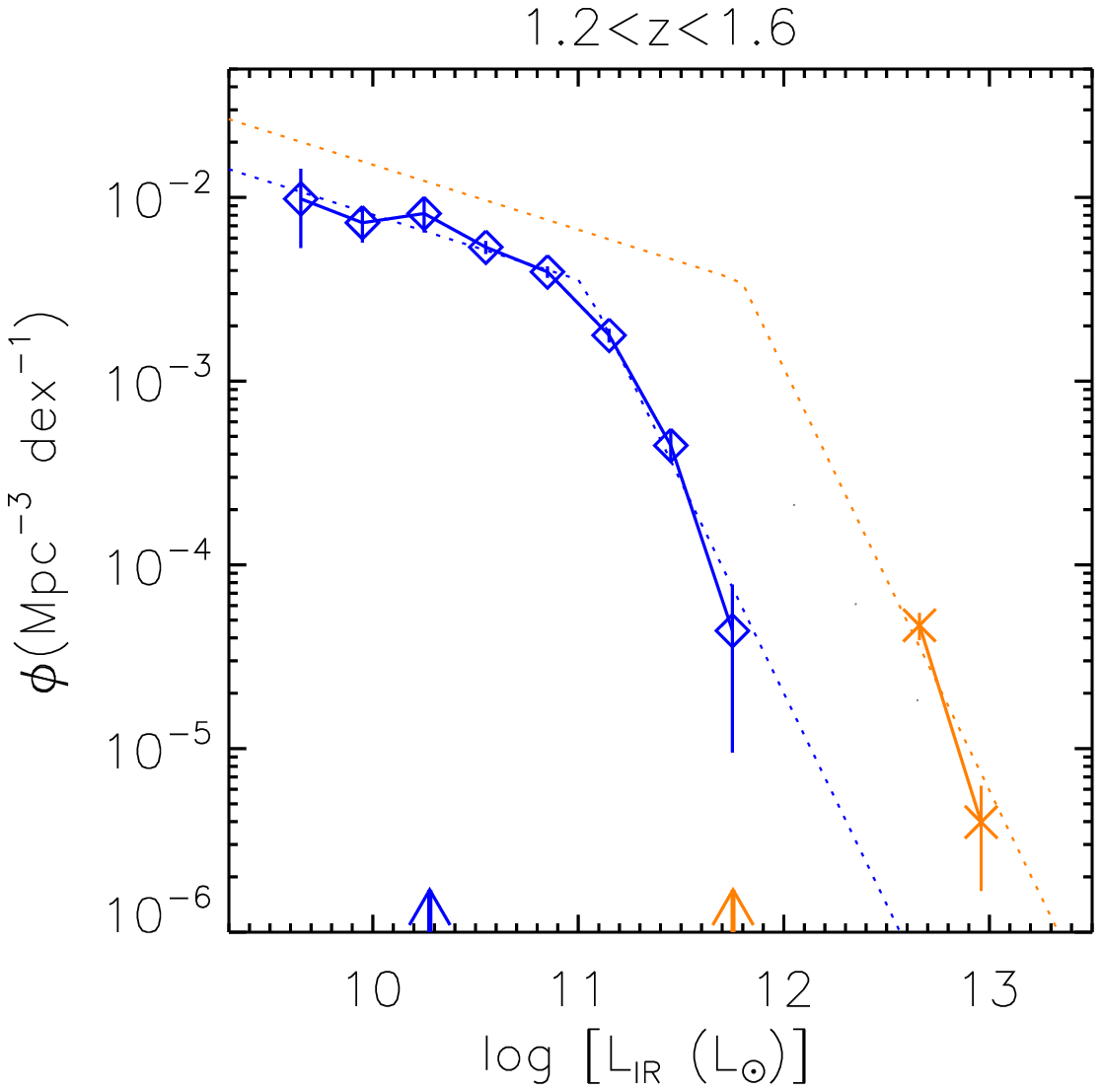}
\end{center}
\caption{
The TIR LFs from the SED fit.
The top panel shows results from all the redshift bins.
The succeeding panels show one redshift bin at a time for clarity.
Vertical arrows show the luminosity corresponding to the flux limit at the central redshift in each redshift bin. 
 We overplot z=0 IR LF based on the AKARI FIR all sky survey in the black dash-dot line \citep{Goto2011SDSS}.
 Overplotted previous studies are taken from 
\citet{2005ApJ...632..169L} in the dark-green, dash-dotted line,
\citet{2009A&A...496...57M} in the dark-red,  dash-dotted line, and 
 \citet{2007ApJ...667L...9H} in the dark-yellow,  dash-dotted line at several redshifts as marked in the figure.
}\label{fig:TIR_LF}
\end{figure*}

 The obtained $L_{\mathrm{TIR}}$ LFs are shown in Fig. \ref{fig:TIR_LF}.
 The uncertainties are estimated through the Monte Carlo simulations ($\S$\ref{sec:vmax}).
 For a local benchmark, we overplot  \citet{Goto2011SDSS}, who cross-correlated the SDSS and the AKARI FIR all sky survey to derive local IR LFs with 2357 galaxies.
Readers are also refereed to Marchetti et al. (submitted) for local LFs based on Herschel HerMES wide fields.
 The TIR LFs show a strong evolution compared to local LFs. 
 At $0.25<z<1.3$, $L^*_{\mathrm{TIR}}$ evolves as $\propto (1+z)^{2.8\pm0.1}$. 



 Using the same methodology as in previous sections, 
 we integrate the $L_{\mathrm{TIR}}$ LFs in Fig. \ref{fig:TIR_LF} through a double-power law fit (eq. \ref{eqn:lumfunc2p} and \ref{eqn:lumfunc2p2}). The resulting total luminosity density ($\Omega_{\mathrm{IR}}$) will be shown in  $\S$\ref{sec:SFH}, along with results from other wavelengths.


\subsection{Evolution of $\Omega_{\mathrm{IR}}$}\label{sec:SFH}

\begin{figure*}
\begin{center}
\includegraphics[scale=.80]{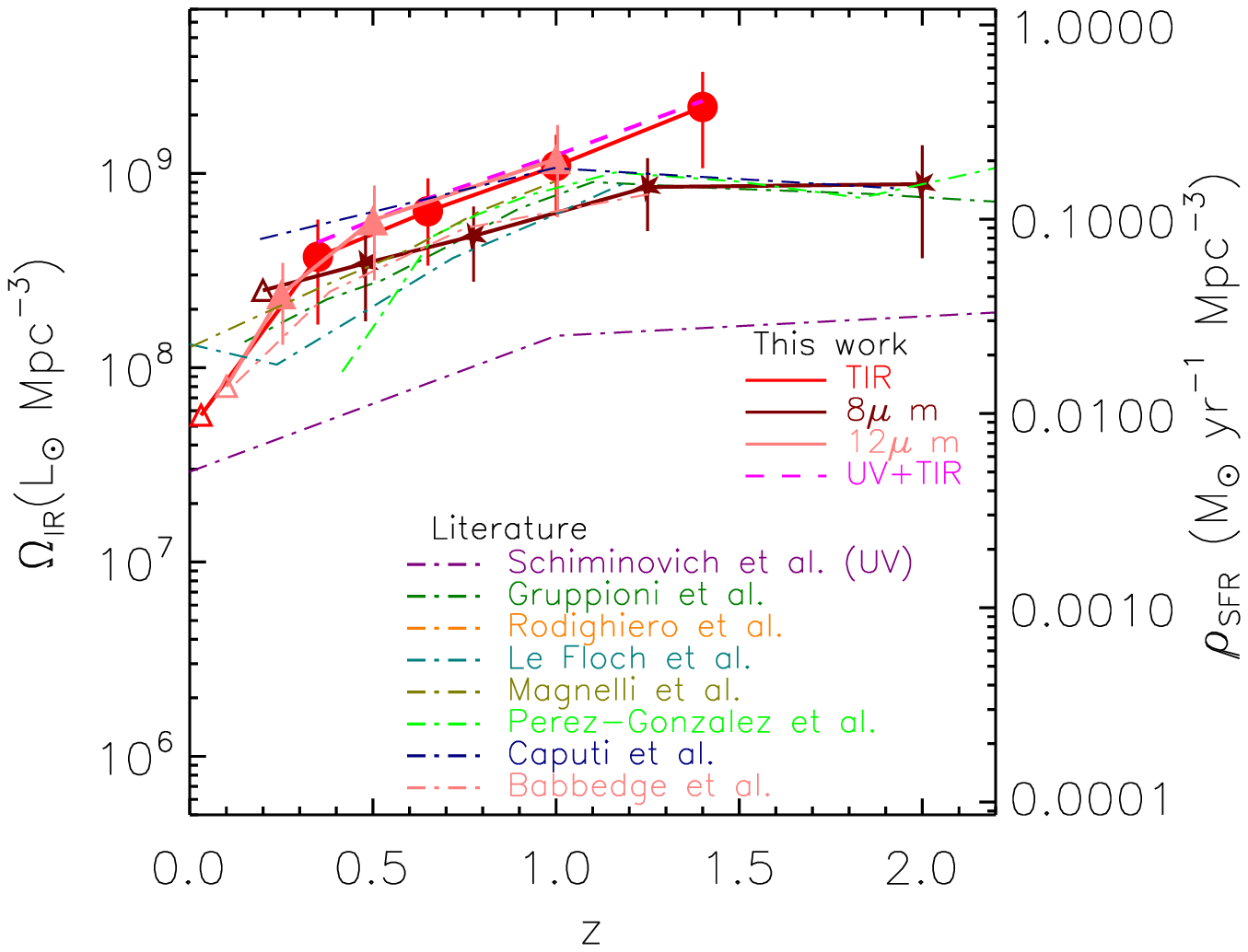}
\end{center}
\caption{
The evolution of the TIR luminosity density based on TIR LFs (red circles,  Table \ref{tab:madau_value}), 8$\mu$m LFs (stars, Table \ref{tab:madau_value_8um}), and 12$\mu$m LFs (filled triangles, Table \ref{tab:madau_value_12um}). 
Values are presented in Table \ref{tab:madau_value}.
Overplotted dot-dashed lines are estimates from the literature: 
\citet{2005ApJ...632..169L}, 
\citet{2009A&A...496...57M}, 
\citet{2005ApJ...630...82P}, 
\citet{2007ApJ...660...97C},   
\citet{2013MNRAS.432...23G},
\citet{2010A&A...515A...8R},
and \citet{2006MNRAS.370.1159B} are in cyan, yellow, green, navy, dark green, orange, and pink, respectively.
The purple dash-dotted line shows the UV estimate by \citet{2005ApJ...619L..47S}.
The pink dashed line shows the total estimate of IR (TIR LF) and UV \citep{2005ApJ...619L..47S}. The open triangles are low-z results from \citet{Goto2011SDSS,2007ApJ...664..840H,2014ApJ...788...45T} in TIR,  8$\mu$m  and 12$\mu$m, respectively. 
}\label{fig:TLD_all}
\end{figure*}

\begin{figure*}
\begin{center}
\includegraphics[scale=.80]{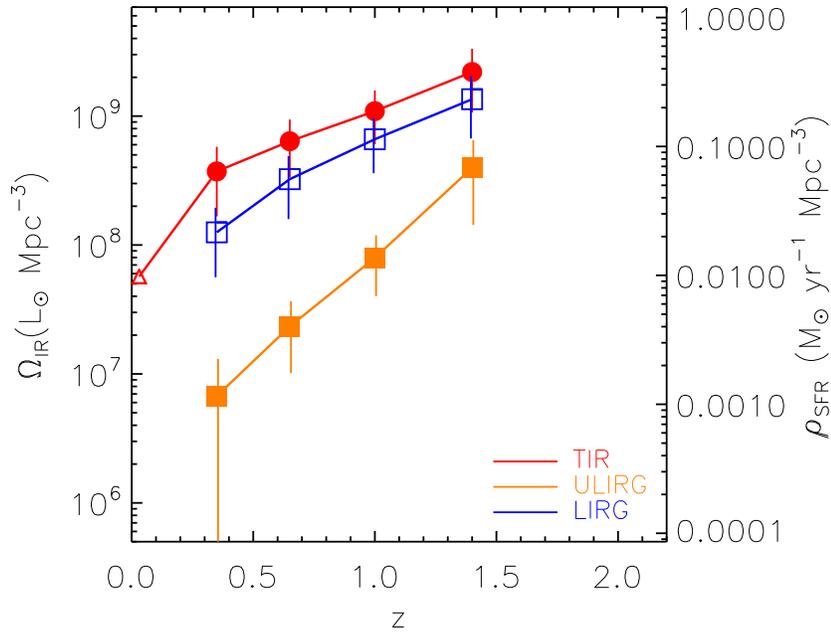}
\end{center}
\caption{
The evolution of the TIR luminosity density based on TIR LFs (red circles,  Table \ref{tab:madau_value}). The blue open squares and orange filled squares corresponds to 
LIRG and ULIRGs samples only, also based on our $L_{\mathrm{TIR}}$ LFs. 
Plotted values are presented in Table \ref{tab:madau_value}.
}\label{fig:TLD_ulirg}
\end{figure*}

\begin{table*}
 \centering
 \begin{minipage}{180mm}
  \caption{TIR luminosity density values and contributions by LIRG and ULIRG as a function redshift as in Fig. \ref{fig:TLD_all}}\label{tab:madau_value}
  \begin{tabular}{@{}ccclllcccc@{}}
  \hline
 z &    $\Omega_{\mathrm{TIR}}(L_\odot Mpc^{-3}/10^8)$ & $\Omega_{LIRG}(L_\odot Mpc^{-3}/10^8)$ & $\Omega_{ULIRG}(L_\odot Mpc^{-3}/10^8)$ \\ 
 \hline
 \hline
0.35 &  3.7$\pm$ 2.1 &  1.25$\pm$0.69 &  0.07$\pm$ 0.06\\
0.65 &  6.4$\pm$ 3.0 &  3.24$\pm$1.65 &  0.23$\pm$ 0.13\\
1.00 & 10.9$\pm$ 4.9 &  6.61$\pm$3.01 &  0.80$\pm$ 0.39\\
1.40 & 22.0$\pm$11.3 & 13.51$\pm$6.82 &  3.97$\pm$ 2.53\\
 \hline
\end{tabular}
\end{minipage}
\end{table*}

\begin{table}
 \centering
 \begin{minipage}{80mm}
  \caption{TIR luminosity density values based on 8$\mu$m LFs,  shown with the brown stars in Fig. \ref{fig:TLD_all}}\label{tab:madau_value_8um}
  \begin{tabular}{@{}ccclllcccc@{}}
  \hline
 z &    $\Omega_{TIR_{8\mu m}}(L_\odot Mpc^{-3}/10^8)$ \\ 
 \hline
 \hline
0.48 &  3.5$\pm$ 1.7\\
0.77 &  4.8$\pm$ 2.0\\
1.25 &  8.5$\pm$ 3.5\\
2.00 &  8.8$\pm$ 5.1\\
 \hline
\end{tabular}
\end{minipage}
\end{table}

\begin{table}
 \centering
 \begin{minipage}{80mm}
  \caption{TIR luminosity density values based on 12$\mu$m LFs, shown with the pink filled-triangles in Fig. \ref{fig:TLD_all}}\label{tab:madau_value_12um}
  \begin{tabular}{@{}ccclllcccc@{}}
  \hline
 z &    $\Omega_{TIR_{12\mu m}}(L_\odot Mpc^{-3}/10^8)$ \\ 
 \hline
 \hline
0.25 &  2.4$\pm$ 1.1\\
0.50 &  5.7$\pm$ 2.9\\
1.00 & 11.8$\pm$ 5.9\\
 \hline
\end{tabular}
\end{minipage}
\end{table}

In this section, we measure the evolution of $\Omega_{\mathrm{IR}}$ as a function of redshift. In Fig. \ref{fig:TLD_all}, we plot $\Omega_{\mathrm{IR}}$ estimated from the TIR LFs (red circles, Table \ref{tab:madau_value}), 8$\mu$m LFs (brown stars, Table \ref{tab:madau_value_8um}), and 12$\mu$m LFs (pink filled triangles, Table \ref{tab:madau_value_12um}), as a function of redshift. Estimates based on 12$\mu$m LFs  and TIR LFs agree with each other very well, while those from 8$\mu$m LFs are slightly lower by a factor of a few than others. However, they are within one sigma of each other. 
Our 8$\mu$m LFs flattens at $z>1.4$. This is in good agreement with \citet{2007ApJ...660...97C} as shown with the dark blue, dash-dotted line.

We compare our results with various papers in the literature \citep{2005ApJ...632..169L,2006MNRAS.370.1159B,2007ApJ...660...97C,2005ApJ...630...82P,2009A&A...496...57M,2013MNRAS.432...23G,2010A&A...515A...8R} in the dash-dotted lines. Our $\Omega_{\mathrm{IR}}$ shows a very good agreement with these at $0<z<1.2$, as almost all of the dash-dotted lines lying within our error bars of  $\Omega_{\mathrm{IR}}$ from $L_{\mathrm{TIR}}$ and 12$\mu$m LFs. This is because an estimate of an integrated value such as $\Omega_{\mathrm{IR}}$ is more reliable than that of LFs.

We parametrize the evolution of $\Omega_{\mathrm{IR}}$ using the following function.

\begin{equation} \label{eqn:SFRevolution}
\Omega_{\mathrm{IR}}(z) \propto (1+z)^{\gamma}
\end{equation}

\noindent By fitting this to the $\Omega_{\mathrm{IR}}$, we obtained 
$\gamma = 4.2\pm 0.4$.  
This is consistent with most previous works. For example, 
\citet{2005ApJ...632..169L} obtained $\gamma = 3.9\pm 0.4$,
\citet{2005ApJ...630...82P} obtained $\gamma = 4.0\pm 0.2$,
\citet{2006MNRAS.370.1159B} obtained $\gamma = 4.5^{+0.7}_{-0.6}$,
\citet{2009A&A...496...57M} obtained $\gamma = 3.6\pm 0.4$.
\citet{2013MNRAS.432...23G,2013A&A...554A..70B} obtained $\gamma = 3.55\pm 0.1$.
\citet{2010A&A...515A...8R} obtained $\gamma = 3.55\pm 0.1$.
The agreement confirms a strong evolution of $\Omega_{\mathrm{IR}}$. 

%


\subsection{Differential evolution among ULIRG and LIRG}

In Fig. \ref{fig:TLD_ulirg}, we plot the contributions to $\Omega_{\mathrm{IR}}$ from LIRGs (Luminous InfraRed Galaxies; $L_{\mathrm{TIR}}>10^{11}L_{\odot}$) 
 and ULIRGs  (Ultra-Luminous InfraRed Galaxies; $L_{\mathrm{TIR}}>10^{12}L_{\odot}$, measured from TIR LFs), with the blue open squares and orange filled squares, respectively. Both LIRGs and ULIRGs show strong evolution.
 From $z$=0.35 to $z$=1.4, $\Omega_{\mathrm{IR}}$ by LIRGs increases by a factor of $\sim$1.8, and 
  $\Omega_{\mathrm{IR}}$ by ULIRGs increases by a factor of $\sim$10.
 The physical origin of ULIRGs in the local Universe is often a merger/interaction  \citep{1996ARA&A..34..749S,1998ApJ...501L.167T,2005MNRAS.360..322G}. It would be interesting to investigate 
 whether the merger rate also increases in proportion to the ULIRG fraction, or if different mechanisms can also produce ULIRGs at higher redshift  \citep[see early work by][]{2010MNRAS.405..234S,2012ApJ...757...23K,2013MNRAS.432.2012T,2013MNRAS.430.1158S}.

\subsection{Comparison to the UV estimate}

\begin{figure}
\begin{center}
\includegraphics[scale=0.6]{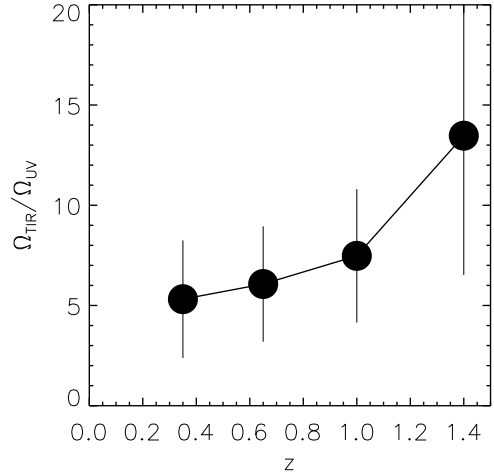}
\end{center}
\caption{
 $\Omega_{\mathrm{TIR}}$ to $\Omega_{\mathrm{UV}}$ ratio shown as a function of redshift.
}\label{fig:UVTIR}
\end{figure}


We have been emphasizing the importance of IR probes of the total SFR density (SFRD) of the Universe. 
However, the IR estimates do not take into account the contribution of the unabsorbed UV light produced by young stars. Therefore, it is important to estimate how significant this direct UV contribution is.
 
\citet{2005ApJ...619L..47S} based on the GALEX data supplemented by the VVDS spectroscopic redshifts, found that the energy density measured at 1500\AA~ evolves as $\propto (1+z)^{2.5\pm0.7}$ at $0<z<1$ and $\propto (1+z)^{0.5\pm0.4}$ at $z>1$.
We overplot their UV estimate of $\rho_{\mathrm{SFR}}$ with the purple dot-dashed line in Fig. \ref{fig:TLD_all}. The UV estimate is almost a factor of 10 smaller than the IR estimate at most of the redshifts, confirming the importance of IR probes when investigating the evolution of the total cosmic star formation density. 
In Fig. \ref{fig:TLD_all} we also plot the total SFD (or $\Omega_{total}$) by adding $\Omega_{\mathrm{UV}}$ and $\Omega_{\mathrm{TIR}}$, with the magenta dashed line. 
 In Fig. \ref{fig:UVTIR}, 
 we show the ratio of the IR-to-UV SFRD ($\Omega_{\mathrm{TIR}}$/ $\Omega_{\mathrm{UV}}$) as a function of redshift. 
Although the errors are large,  Fig. \ref{fig:UVTIR} agrees with \citet{2005A&A...440L..17T} and recent work based on Herschel by \citet{2013A&A...554A..70B}.
Our result suggests that $\Omega_{\mathrm{TIR}}$/ $\Omega_{\mathrm{UV}}$ evolves strongly with redshift: from 5.3 at z=0.35 to 13.5 at z=1.4, i.e., $\Omega_{\mathrm{TIR}}$ explains already 85\% of $\Omega_{total}(=\Omega_{\mathrm{UV}}+\Omega_{\mathrm{TIR}})$ at $z$=0.25, and that by $z$=1.4, 93\% of the cosmic SFD is explained by the infrared. This implies that $\Omega_{\mathrm{TIR}}$ provides a good approximation of the  $\Omega_{total}$ at $z>1$.
 
\section{Discussion}
\label{091258_25May15}
 In this work, we have expanded our previous work on mid-IR LFs to the entire AKARI NEP deep field, using our new wide coverage CFHT optical/near-IR photometry. Our LFs are in good agreement with our previous work based on a smaller (40\%) area \citep{GotoTakagi2010}, but with much reduced errors. Our previous work had notable shot noise, which made it difficult to discuss evolutionary and environmental effects \citep{cluster_LF}. The LFs in this work are more stable, and allow us more precise discussion of the redshift evolution.
This highlights the importance of wide field coverage, not only to reduce statistical errors, but also to overcome cosmic variance.
AKARI has also obtained shallower but wider NEP wide-field data, which covered 5.4 deg$^2$ \citep{2009PASJ...61..375L,2014ApJS..214...20J,2012A&A...548A..29K}. We are making an effort to probe the bright-end of the LFs from these data, using spectroscopic redshifts \citep{Kimetal}. 
In parallel, we are trying to obtain accurate photometric redshifts over the NEP-wide field. Initial data in $r$-band have just been taken using the new Subaru Hyper-Suprime Cam ($r\sim$27mag, PI: Goto).  

 By combining Herschel/SPIRE far-IR data, more accurate estimate of $L_{\mathrm{TIR}}$ can be obtained than the mid-IR only estimation \citep[e.g.,][]{2014ApJ...784..137K}. In the AKARI NEP deep field, there exists Herschel data (PI: Sergeant). 
  However, SPIRE data in the field is relatively shallow. The SPIRE source density is only 10\% of that of AKARI sources. Only 1.1\% of AKARI sources had a counterpart in SPIRE in a simple positional match within 5''. Therefore, if we want to exploit AKARI's data to the faintest flux limit, we need to proceed without SPIRE detection for 99\% of the sources. Alternately, if we limit our sample to SPIRE detected sources, we can only use 1.1\% of the AKARI sources. 
One of the main purposes of obtaining the new CFHT imaging data was to increase the statistics. Therefore, in this paper, we decided not to limit our sample to Herschel detected sources. 

 In other deeper fields, however, Herschel has revealed the IR LFs to z=4 \citep{2013MNRAS.432...23G,2013A&A...554A..70B}. This achievement has changed the role of the mid-IR studies; before Herschel, the mid-IR analysis was an effective way to estimate SFR in the high-z Universe, even considering the large uncertainty in converting mid-IR photometry to $L_{\mathrm{TIR}}$ or SFR. However, now that Herschel has provided direct far-IR measurements, the importance of mid-IR data has shifted toward understanding more details of PAH or hot-dust emission, in comparison with far-IR data, in order to understand the SED evolution between the mid-IR and far-IR. Many such attempts have begun  \citep{2011A&A...533A.119E,2012ApJ...745..182N,2012A&A...545A.141B,2014A&A...566A.136M}. This work provides an important step from the  mid-IR side, based on AKARI's unique continuous 9-band mid-IR coverage over a wide area (0.6 deg$^2$).

\section{Summary}

 We have used recently obtained wide-field optical/near-IR imaging from CFHT. These data covered the entire AKARI NEP deep field, allowing us to utilize, for the first time, all of the infrared  data AKARI has taken in the field.

 AKARI's advantage is its continuous filter coverage in the mid-IR wavelengths (2.4, 3.2, 4.1, 7, 9, 11, 15, 18, and 24$\mu$m). No other telescope including Spitzer and WISE has such a filter coverage in the mid-infrared. The data allowed us to estimate mid-IR luminosity without a large extrapolation based on SED models, which was the largest uncertainty in previous studies. Even for $L_{\mathrm{TIR}}$, the SED fitting is more reliable due to this continuous coverage of mid-IR wavelengths. 

 We have estimated restframe 8$\mu$m, 12$\mu$m, and total infrared luminosity functions, as functions of redshift.
All LFs are in agreement once converted to a TIR LF, and show strong evolution from $z$=0.35 to $z$=2.2.
 $\Omega_{\mathrm{IR}}$ evolves as 
 $\propto (1+z)^{4.2\pm 0.4}$.
We found that the ULIRG (LIRG) contribution increases by a factor of 10 (1.8) from $z$=0.35 to $z$=1.4, suggesting that IR galaxies dominate $\Omega_{\mathrm{IR}}$ at higher redshift.
We estimated that  $\Omega_{\mathrm{IR}}$ includes 85\%  of the cosmic star formation at redshifts less than 1, and more than 90\% at higher redshifts.  
These results are consistent with our previous work \citep{GotoTakagi2010}, but with much reduced statistical errors thanks to the large area coverage of the CFHT and AKARI data. 

\section*{Acknowledgments}

We thank the anonymous referee for many insightful comments, which significantly improved the paper.
We are grateful to K. Murata, T. Miyaji, and F. Mazyed for useful discussion.
 TG acknowledges the support by the Ministry of Science and Technology of Taiwan through grant NSC 103-2112-M-007-002-MY3.
MI acknowledges the support by the National Research Foundation of Korea (NRF) grant, No. 2008-0060544, funded by the Korea government (MSIP).
KM have been supported by the National Science Centre (grant UMO-2013/09/D/ST9/04030).
This work was supported by the National Research Foundation of Korea (NRF) grant funded by the Korea government (MEST; No. 2012-006087).
This research is based on the observations with AKARI, a JAXA project with the participation of ESA.

\label{lastpage}



%
%


\bibliography{201404_goto} 
\bibliographystyle{mnras}

\label{lastpage}

\end{document}